\documentclass[12pt,prd,tightenlines,nofootinbib]{revtex4}
\usepackage{bm}
\usepackage{graphics}
\usepackage{rotating}
\usepackage{epsfig}
\begin{document}
\title{
Semileptonic decays of $D$ and $D_s$ mesons in the relativistic quark model}
\author{R. N. Faustov}\email{faustov@ccas.ru}
\author{V. O. Galkin}\email{galkin@ccas.ru}

\affiliation{Institute of Cybernetics and Informatics in Education, FRC CSC RAS,
  Vavilov Street 40, 119333 Moscow, Russia}

\author{Xian-Wei Kang}\email{11112018023@bnu.edu.cn}
\affiliation{College of Nuclear Science and Technology, Beijing Normal University, Beijing 100875, China}
\begin{abstract}
The form factors parameterizing the weak
$D$ and $D_s$ transitions to light pseudoscalar and vector mesons are calculated in the framework of the
relativistic quark model based on the quasipotential approach. The
special attention is paid to the systematic account of the
relativistic effects including transformation of the meson wave
function from the rest to  moving reference frame and contributions of
the intermediate negative-energy states. The form factors are
expressed through the overlap integrals of the meson wave functions,
which are taken from previous studies of meson spectroscopy. They are
calculated in the whole range of the transferred momentum
$q^2$. Convenient parameterization of the form factors which
accurately reproduces numerical results is given. The obtained values
of the form factors and their ratios at $q^2=0$ agree well with the
ones extracted form the experimental data. On
the basis of these form factors and helicity formalism, differential and total
semileptonic decay rates of $D$ and $D_s$ mesons as well as different asymmetries and
polarization parameters are calculated. The detailed comparison of the
obtained results with other theoretical calculations and experimental
data is given.
\end{abstract}

\maketitle

\section{Introduction}
\label{sec:intr}

Semileptonic decays of heavy mesons provide an important information on the
values of the Cabbibo-Kobayashi-Maskawa (CKM) matrix elements
$V_{Qq}$ (with $Q$ denoting the heavy quark and $q$ the light one),  which are essential ingredients of the standard
model. Experimentally such decays can be measured more accurately than
pure leptonic ones since there is no helicity suppression for
them. Theoretically semileptonic decays are significantly less
complicated than hadronic ones as they  contain one meson and a
lepton pair in the final state. The lepton part is easily calculated
using standard methods, while the hadronic part factorizes
thus reducing theoretical uncertainties. The hadronic matrix element
is usually parameterized by the set of invariant form factors, which
are calculated using nonperturbative approaches based on quantum
chromodynamics (QCD), such as lattice QCD, QCD sum rules, potential quark models.

Recently significant experimental progress has been achieved in
studying semileptonic decays of the open charm mesons \cite{pdg}. More precise and detailed
measurements of the absolute and differential branching fractions and
form factors for $D$ and $D_s$ decays to pseudoscalar and vector
mesons became available due to high statistics accumulated at BES III
\cite{besiii-1,besiii-2,besiii-3,besiii-4,besiii-5,besiii-6,besiii-7,besiii-8,besiii-9}. Various CKM- favored and suppressed decay modes both
with positron and muon were investigated. This allows one to check the
lepton universality in $D$ meson decays. Note that possible hints of
its violation were recently found in $B$ decays \cite{lfv}. More
precise and comprehensive data are expected form BES III and Belle
II \cite{belleii} in near future.

In this paper we calculate the matrix elements of the flavor changing
charged weak current between initial $D$ or $D_s$ mesons and final light
pseudoscalar or vector mesons in the framework of the relativistic quark
model based on the quasipotential approach. This model was successfully
applied for the calculations of the hadron spectroscopy
\cite{efg-1,efg-2,efg-3,efg-4} and weak
decays \cite{efg-5,efg-6,efg-7,fg-1,fg-2}. It was found that relativistic effects play very important
role both for light and heavy hadrons. Thus the form factors are
calculated with the consistent account of the relativistic
quark dynamics. They are expressed through the overlap integrals of the meson
wave functions which are known from the study of their
spectroscopy. The momentum transfer $q^2$ dependence of form factors
is explicitly determined in the whole kinematical range without
additional assumptions and extrapolations. Then we use
these form factors and the helicity formalism for the calculation of
the differential and total branching fractions as well as polarization
and asymmetry parameters. We also compare our results with
available experimental data and previous predictions.

The paper is organized as follows. In Sec.~\ref{sec:rqm} we briefly
describe our relativistic quark model with special emphasis on
calculation of the weak decay matrix elements between meson states
with the account of relativistic effects. This model is applied in
Sec.~\ref{sec:dff} to the consideration of semileptonic decay form
factors of open charm mesons. We give the analytic expressions for the
form factors which
accurately reproduce the numerical results for the momentum transfer
$q^2$ dependence of the form factors in the whole accessible kinematical range
and compare them with available data. Then in Sec.~\ref{sec:sd} we use
these form factors to calculate  the differential and total
$D$ and $D_s$ meson semileptonic decay rates and different asymmetries
and polarization parameters. Decays both with positrons and muons are
considered. This allows us to give predictions for the ratios of the
corresponding  decay rates which can be used for the test of the lepton
universality in charm meson decays. Finally, Sec.~\ref{sec:conc}
contains our conclusions.

\section{Relativistic quark model}
\label{sec:rqm}

For the calculation of meson properties we employ the relativistic
quark model based on the quasipotential approach. In this model a
meson with the mass  $M$ is described by the wave function $\Psi_{M}({\bf p})$ of the quark-antiquark bound
state which satisfies the Schr\"odinger-like quasipotential equation
\cite{efg-1}
\begin{equation}
\label{eq:quas}
{\left(\frac{b^2(M)}{2\mu_{R}}-\frac{{\bf
p}^2}{2\mu_{R}}\right)\Psi_{M}({\bf p})} =\int\frac{d^3 q}{(2\pi)^3}
 V({\bf p,q};M)\Psi_{M}({\bf q}),
\end{equation}
where $m_{1,2}$ are the quark masses, ${\bf p}$ is the relative quark
momentum.  The relative momentum squared in the center of mass system
on the mass shell is given by
\begin{equation}
{b^2(M) }
=\frac{[M^2-(m_1+m_2)^2][M^2-(m_1-m_2)^2]}{4M^2},
\end{equation}
and the relativistic reduced mass is defined by
\begin{equation}
\mu_{R}=\frac{M^4-(m^2_1-m^2_2)^2}{4M^3}.
\end{equation}
The kernel of this equation $V({\bf p,q};M)$ is the QCD-motivated
quark-antiquark potential which is constructed by the off-mass-shell
scattering amplitude projected on the positive energy states. We
assume \cite{efg-1} that it consists from the one-gluon exchange term which
dominates at small distances and a mixture of the scalar and vector
linear confining interactions which dominate at large
distances. Moreover, we assume that the long-range vertex of the
confining vector interaction contains additional Pauli term. Then the
quasipotential is given by
 \begin{equation}
\label{eq:qpot}
V({\bf p,q};M)=\bar{u}_1(p)\bar{u}_2(-p){\mathcal V}({\bf p}, {\bf
q};M)u_1(q)u_2(-q),
\end{equation}
with
$${\mathcal V}({\bf p},{\bf q};M)=\frac{4}{3}\alpha_sD_{ \mu\nu}({\bf
k})\gamma_1^{\mu}\gamma_2^{\nu}
+V^V_{\rm conf}({\bf k})\Gamma_1^{\mu}({\bf k})
\Gamma_{2;\mu}({\bf k})+V^S_{\rm conf}({\bf k}),\qquad {\bf k=p-q},$$
where $\alpha_s$ is the QCD coupling constant, $D_{\mu\nu}$ is the
gluon propagator in the Coulomb gauge, and $\gamma_{\mu}$ and $u(p)$ are the Dirac matrices and
spinors, respectively. The long-range vector vertex has the form
\begin{equation}
\label{eq:pauli}
\Gamma_{\mu}({\bf k})=\gamma_{\mu}+ \frac{i\kappa}{2m}\sigma_{\mu\nu}k^{\nu},
\end{equation}
where $\kappa$ is the long-range anomalous chromomagnetic quark
moment. In the nonrelativistic limit confining vector and scalar
potentials reduce to
\begin{equation}
  \label{eq:conf}
V_{\rm conf}^V(r)=(1-\varepsilon)(Ar+B),\qquad
V_{\rm conf}^S(r) =\varepsilon (Ar+B),
\end{equation}
and in the sum they reproduce the linear rising potential
\begin{equation}
\label{nr}
V_{\rm conf}(r)=V_{\rm conf}^S(r)+V_{\rm conf}^V(r)=Ar+B,
\end{equation}
where $\varepsilon$ is the mixing coefficient. Thus this quasipotential
can be viewed as the relativistic generalization of the
nonrelativistic Cornell potential
\begin{equation}
  \label{eq:cornell}
  V_{\rm NR}(r)=-\frac43\frac{\alpha_s}r +Ar +B.
\end{equation}
Our quasipotential contains both spin-independent and spin-dependent relativistic
contributions.

All parameters of the model were fixed from the previous consideration
of hadron spectroscopy and decays \cite{efg-1}. Thus the values of constituent quark masses
are $m_b=4.88$ GeV, $m_c=1.55$ GeV, $m_s=0.5$ GeV, $m_{u,d}=0.33$ GeV;
the parameters of the linear potential are $A=0.18$ GeV$^2$ and
$B=-0.30$ GeV; the mixing parameter of the vector and scalar confining
potential is $\varepsilon=-1$, while the anomalous chromomagnetic
quark moment $\kappa=-1$. We take the running QCD coupling constant
with infrared freezing
\begin{equation}
  \label{eq:alpha}
\alpha_s(\mu)=\frac{4\pi}{\beta_0\ln\frac{\mu^2+M_0^2}{\Lambda^2}},
\end{equation}
where $\beta_0=11-\frac23n_f$, $n_f$ is the number of flavors, $\Lambda=413$~MeV,
$M_0=2.24\sqrt{A}=0.95$~GeV and the scale $\mu$ is set to
$\frac{2m_1m_2}{m_1+m_2}$.

The spectroscopy of heavy-light and light mesons was discussed in
detail in Refs.~ \cite{efg-3,efg-2}. The calculated masses for both of the
ground and excited states were found in agreement with available
experimental data and exhibit linear Regge trajectories. The meson wave functions were also calculated and
can be used for the evaluation of the meson decays.

For the consideration of the $D$ meson semileptonic decays it is
necessary to calculate the hadronic matrix element of the local
current governing the $c\to f$ ($f=s,d$) weak transition. In
the quasipotential approach the matrix element of this weak current
$J_\mu^W=\bar f\gamma_\mu(1-\gamma_5)c$
between the initial $D_{(s)}$ meson with four-momentum $p_{D_{(s)}}$ and final
meson $F$ with four-momentum $p_F$ is given by \cite{efg-5}
\begin{equation}\label{eq:mel}
\langle F(p_F) \vert J^W_\mu \vert D_{(s)}(p_{D_{(s)}})\rangle
=\int \frac{d^3p\, d^3q}{(2\pi )^6} \bar \Psi_{{F}\,{\bf p}_F}({\bf
p})\Gamma _\mu ({\bf p},{\bf q})\Psi_{D_{(s)}\,{\bf p}_{D_{(s)}}}({\bf q}),
\end{equation}
where $\Psi_{M\,{\bf p}_M}$ are the initial and final meson wave
functions projected on the positive energy states and boosted to the
moving reference frame with the three-momentum ${\bf p}_M$. The vertex
function
\begin{equation}
  \Gamma=\Gamma^{(1)}+\Gamma^{(2)},\label{eq:G}
\end{equation}
 where $\Gamma^{(1)}$ is
the leading-order vertex function which corresponds to the impulse
approximation
\begin{equation} \label{eq:G1}
\Gamma_\mu^{(1)}({\bf
p},{\bf q})=\bar u_{f}(p_f)\gamma_\mu(1-\gamma^5)u_c(q_c)
(2\pi)^3\delta({\bf p}_{q}-{\bf q}_{q})\end{equation}
and contains the $\delta$ function responsible for the momentum
conservation on the spectator $q$ antiquark line. The vertex function
$\Gamma^{(2)}$ takes into account interaction of the active quarks ($c,f$) with
the spectator antiquark ($q$) and includes the negative-energy part of
the active quark propagator. It is the consequence of the projection on
the positive energy states and  has the form
\begin{eqnarray}
  \label{eq:G2}
\Gamma_\mu^{(2)}({\bf
p},{\bf q})&=&\bar u_f(p_f)\bar u_q(p_{q}) \Bigl\{{\cal V}({\bf p}_{q}-{\bf
q}_{q})\frac{\Lambda_{f}^{(-)}(k')}{ \epsilon_{f}(k')+
\epsilon_{f}(q_c)}\gamma_1^0 \gamma_{1\mu}(1-\gamma_1^5)\nonumber \\
& &+\gamma_{1\mu}(1-\gamma_1^5)
\frac{\Lambda_c^{(-)}(k)}{\epsilon_c(k)+\epsilon_c(p_f)}\gamma_1^0
{\cal V}({\bf p}_{q}-{\bf
q}_{q})\Bigr\}u_c(q_c) u_q(q_{q}),
\end{eqnarray}
where ${\bf k}={\bf p}_f-{\bf\Delta};\
{\bf k}'={\bf q}_c+{\bf\Delta};\ {\bf\Delta}={\bf
p}_{F}-{\bf p}_{D}$; $\epsilon(p)=\sqrt{{\bf p}^2+m^2}$; and the projection operator on the
negative-energy states
$$\Lambda^{(-)}(p)=\frac{\epsilon(p)-\bigl( m\gamma
  ^0+\gamma^0({\bm{ \gamma}{\bf p}})\bigr)}{ 2\epsilon (p)}.$$
Note that the $\delta$ function in the vertex function $\Gamma^{(1)}$
[Eq.~(\ref{eq:G1})] allows us to take off one of the integrals in the
expression for the matrix element Eq.~(\ref{eq:mel}). As the result the
usual expression for the matrix element as the overlap integral of the
meson wave functions is obtained. The contribution $\Gamma^{(2)}$
[Eq.~(\ref{eq:G2})] is significantly more complicated and contains the
quasipotential of the quark-antiquark interaction ${\cal V}$
[Eq.~(\ref{eq:qpot})] which has nontrivial Lorentz-structure. However, it is
possible to use the quasipotential equation (\ref{eq:quas}) to get rid
of one of the integrations in Eq.~(\ref{eq:mel}) and thus get again the
usual structure of the matrix element as the overlap integral of meson
wave functions (for details see Refs.~\cite{efg-5,efg-6}).

Calculations of hadron decays are usually done in the
rest frame of the decaying hadron, the $D_{(s)}$
meson in the considered case,  where the decaying meson momentum ${\bf p}_{D}=0$. Then
the final  meson $F$ is moving with the recoil momentum ${\bf
  \Delta}={\bf p}_F$
and its wave function should be boosted to the moving reference frame.
The wave function of the moving  meson $\Psi_{F\,{\bf\Delta}}$ is connected
with the  wave function in the rest frame
$\Psi_{F\,{\bf 0}}$ by the transformation \cite{efg-5}
\begin{equation}
\label{wig}
\Psi_{F\,{\bf\Delta}}({\bf
p})=D_f^{1/2}(R_{L_{\bf\Delta}}^W)D_q^{1/2}(R_{L_{
\bf\Delta}}^W)\Psi_{F\,{\bf 0}}({\bf p}),
\end{equation}
where $R^W$ is the Wigner rotation, $L_{\bf\Delta}$ is the Lorentz boost
from the meson rest frame to a moving one and $D^{1/2}(R)$ is
the spin rotation matrix.

\section{Weak decay form factors}
\label{sec:dff}

In the standard model  the semileptonic $D$ and $D_s$ meson decays to
a pseudoscalar ($P$) or a vector ($V$) mesons are governed by the
flavor-changing $c\to q\ell\nu_\ell$ ($q=s,d$) current.  The corresponding matrix
element ${\cal M}$ between meson states factorizes in the product of the leptonic ($L_\mu$)
current and the matrix element of the hadronic
($H^\mu$) current with the corresponding CKM matrix element $V_{cq}$
 and the Fermi constant $G_F$
\begin{equation}
  \label{eq:sdme}
  {\cal M}(D_{(s)}\to P(V) \ell\nu_\ell)=\frac{G_F}{\sqrt2}V_{cq}H^\mu L_\mu,
\end{equation}
where $L_\mu=\bar\nu_\ell\gamma_\mu(1-\gamma_5)\ell$ and
$H^\mu=\langle P(V)|\bar q\gamma_\mu(1-\gamma_5)c|D_{(s)}\rangle$. The
leptonic part is easily calculated using the lepton spinors and has a
simple structure, while the hadronic part is significantly more
complicated and requires nonperturbative treatment within QCD.

The hadronic  matrix element of weak current $J^W$ between meson
states is usually
parameterized by the following set of the invariant form factors.
\begin{itemize}
\item  For $D_{(s)}$ transitions to pseudoscalar $P$
  ($\pi,K,\eta,\eta'$) mesons
\begin{eqnarray}
  \label{eq:pff}
 \!\!\!\!\!\!\!\!\!\!\!\!\!\!\! \langle P(p_{P})|\bar q \gamma^\mu c|D_{(s)}(p_{D_{(s)}})\rangle
 & =&f_+(q^2)\left[p_{D_{(s)}}^\mu+ p_{P}^\mu-
\frac{M_{D_{(s)}}^2-M_{P}^2}{q^2}\ q^\mu\right]+
  f_0(q^2)\frac{M_{D_{(s)}}^2-M_{P}^2}{q^2}\ q^\mu,\cr\cr
 \!\!\!\!\!\!\!\!\!\!\!\!\!\!\!\!\langle P(p_{P})|\bar q \gamma^\mu\gamma_5 c|D_{(s)}(p_{D_{(s)}})\rangle
  &=&0,
\end{eqnarray}
\item  For $D_{(s)}$ transitions to vector $V$
  ($\rho,\omega,K^*,\phi$) mesons
\begin{eqnarray}
  \label{eq:vff}
\!\!\!\!\!\!\!\!\!\!\!\!\!\!\!  \langle V(p_{V})|\bar q \gamma^\mu c|D_{(s)}(p_{D_{(s)}})\rangle&=
  &\frac{2iV(q^2)}{M_{D_{(s)}}+M_{V}} \epsilon^{\mu\nu\rho\sigma}\epsilon^*_\nu
  p_{D_{(s)}\rho} p_{V\sigma},\cr \cr
\label{eq:vff2}
\!\!\!\!\!\!\!\!\!\!\!\!\!\!\!\langle V(p_{V})|\bar q \gamma^\mu\gamma_5 c|D_{(s)}(p_{D_{(s)}})\rangle&=&2M_{V}
A_0(q^2)\frac{\epsilon^*\cdot q}{q^2}\ q^\mu
 +(M_{D_{(s)}}+M_{V})A_1(q^2)\left(\epsilon^{*\mu}-\frac{\epsilon^*\cdot
    q}{q^2}\ q^\mu\right)\cr\cr
&&-A_2(q^2)\frac{\epsilon^*\cdot q}{M_{D_{(s)}}+M_{V}}\left[p_{D_{(s)}}^\mu+
  p_{V}^\mu-\frac{M_{D_{(s)}}^2-M_{V}^2}{q^2}\ q^\mu\right].
\end{eqnarray}
\end{itemize}
At the maximum recoil point ($q^2=0$) these form
factors satisfy the following conditions:
\[f_+(0)=f_0(0),\]
\[A_0(0)=\frac{M_{D_{(s)}}+M_{V}}{2M_{V}}A_1(0)
  -\frac{M_{D_{(s)}}-M_{V}}{2M_{V}}A_2(0).\]

We use the quasipotential approach and the relativistic quark model
discussed in Sec.~\ref{sec:rqm} for the calculation of the weak decay
matrix elements and transition form factors. We substitute the leading
$\Gamma^{(1)}$ [Eq.~(\ref{eq:G1})] and subleading $\Gamma^{(2)}$
[Eq.~(\ref{eq:G2})] vertex functions  in the expression for the matrix
element of the weak current between meson states (\ref{eq:mel}). This
matrix element is considered in the rest
frame of the decaying $D_{(s)}$ meson, then the boost of the final
meson wave function $\Psi_F$ from the rest to  moving reference frame
with the recoil momentum ${\bf\Delta}={\bf p}_F$ should be considered. It is given by Eq.~(\ref{wig}).  Thus
we take into account all relativistic effects including  the
relativistic contributions of intermediate negative-energy states and
relativistic transformations of the meson wave functions. The
resulting expressions for the decay form factors have the form of the
overlap integrals of initial and final meson wave functions. They are
rather cumbersome and are given in Refs.~\cite{efg-5,efg-6}. For the
numerical evaluation of the decay form factors we use the meson wave
functions obtained in calculating their mass spectra
\cite{efg-2,efg-4}. This is a significant advantage of our approach since
in most of the previous model calculations some phenomenological wave
functions (such as Gaussian) were used. Moreover, our relativistic
approach allows us to determine the form factor dependence on the
transferred momentum $q^2$ in the whole accessible kinematical range
without additional approximations and extrapolations.

We find that the numerical results for these decay form factors can
be approximated with high accuracy by the following expressions:

(a) $f_+(q^2),V(q^2),A_0(q^2)$
\begin{equation}
  \label{fitfv}
  F(q^2)=\frac{F(0)}{\displaystyle\left(1-\frac{q^2}{ M^2}\right)
    \left(1-\sigma_1
      \frac{q^2}{M_{D_{(s)}^*}^2}+ \sigma_2\frac{q^4}{M_{D_{(s)}^*}^4}\right)},
\end{equation}

\begin{figure}
\centering
  \includegraphics[width=8cm]{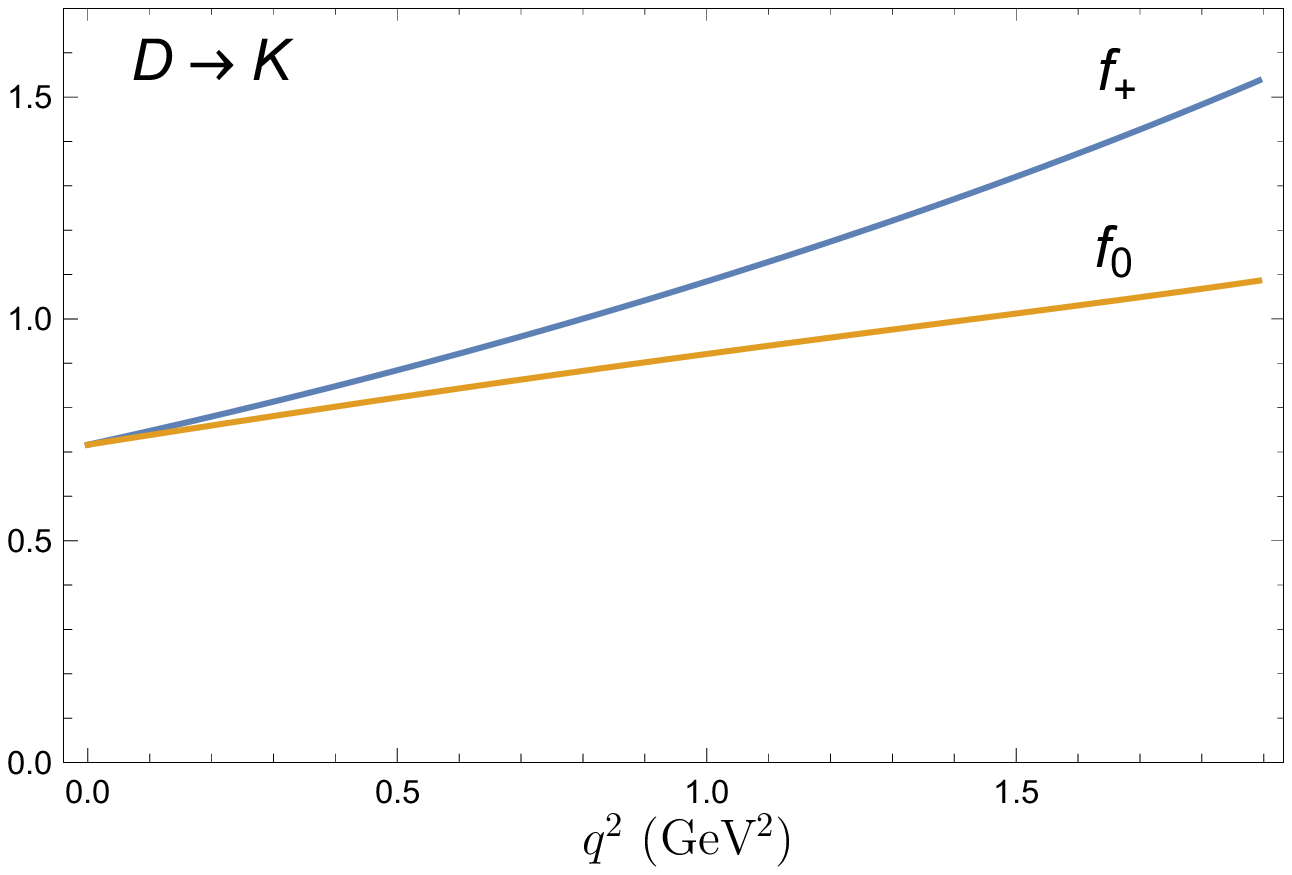}\ \
  \ \includegraphics[width=8cm]{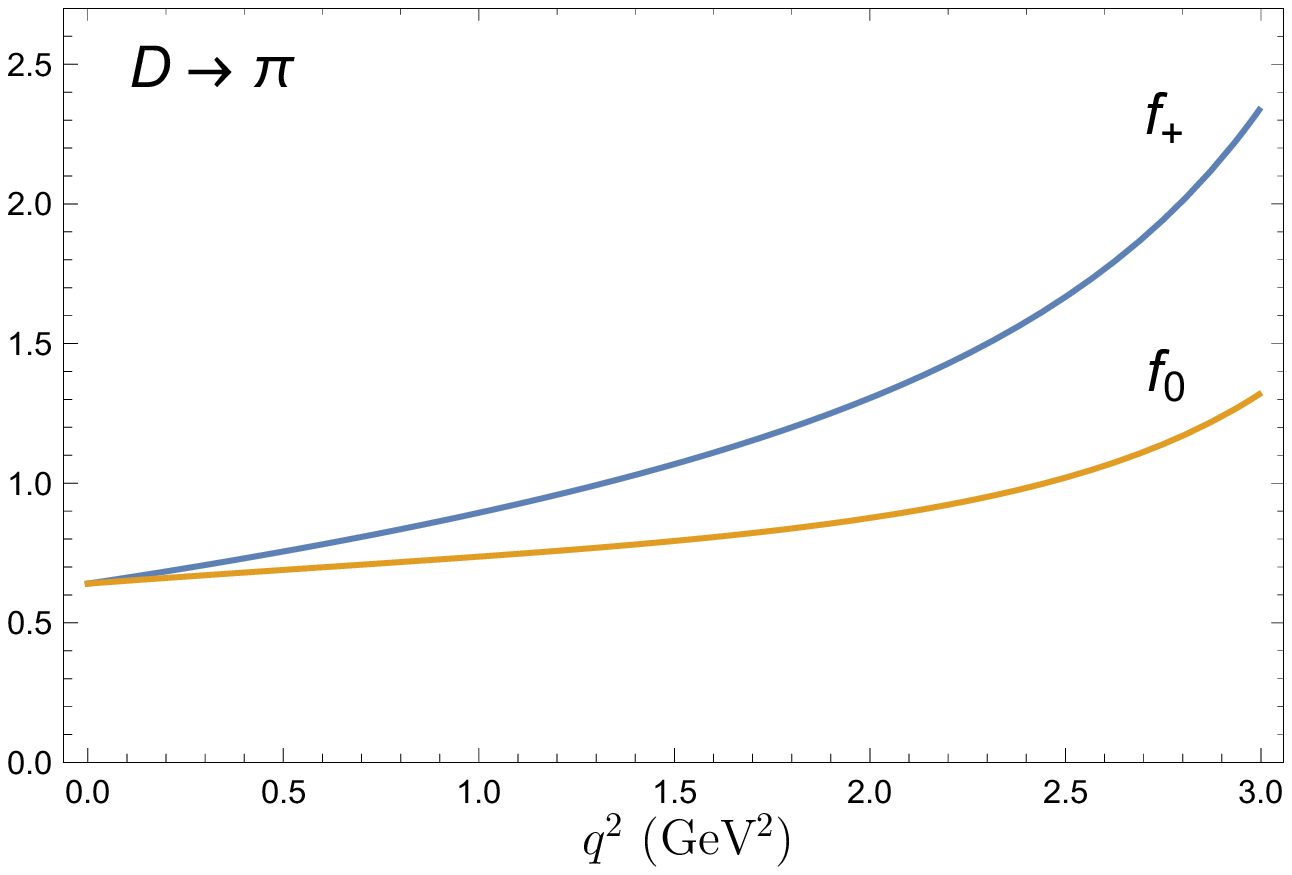}\\
  \includegraphics[width=8cm]{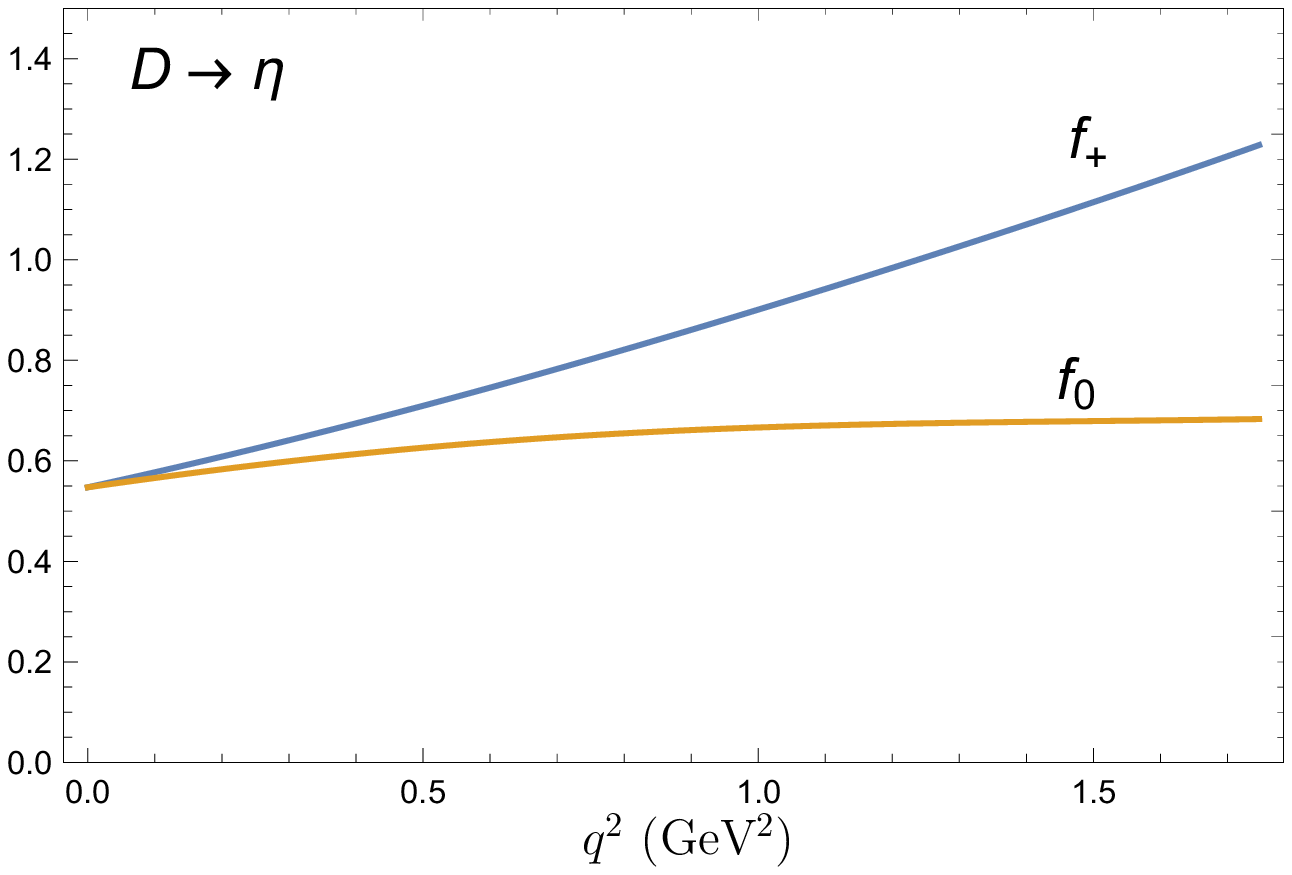}\ \
  \ \includegraphics[width=8cm]{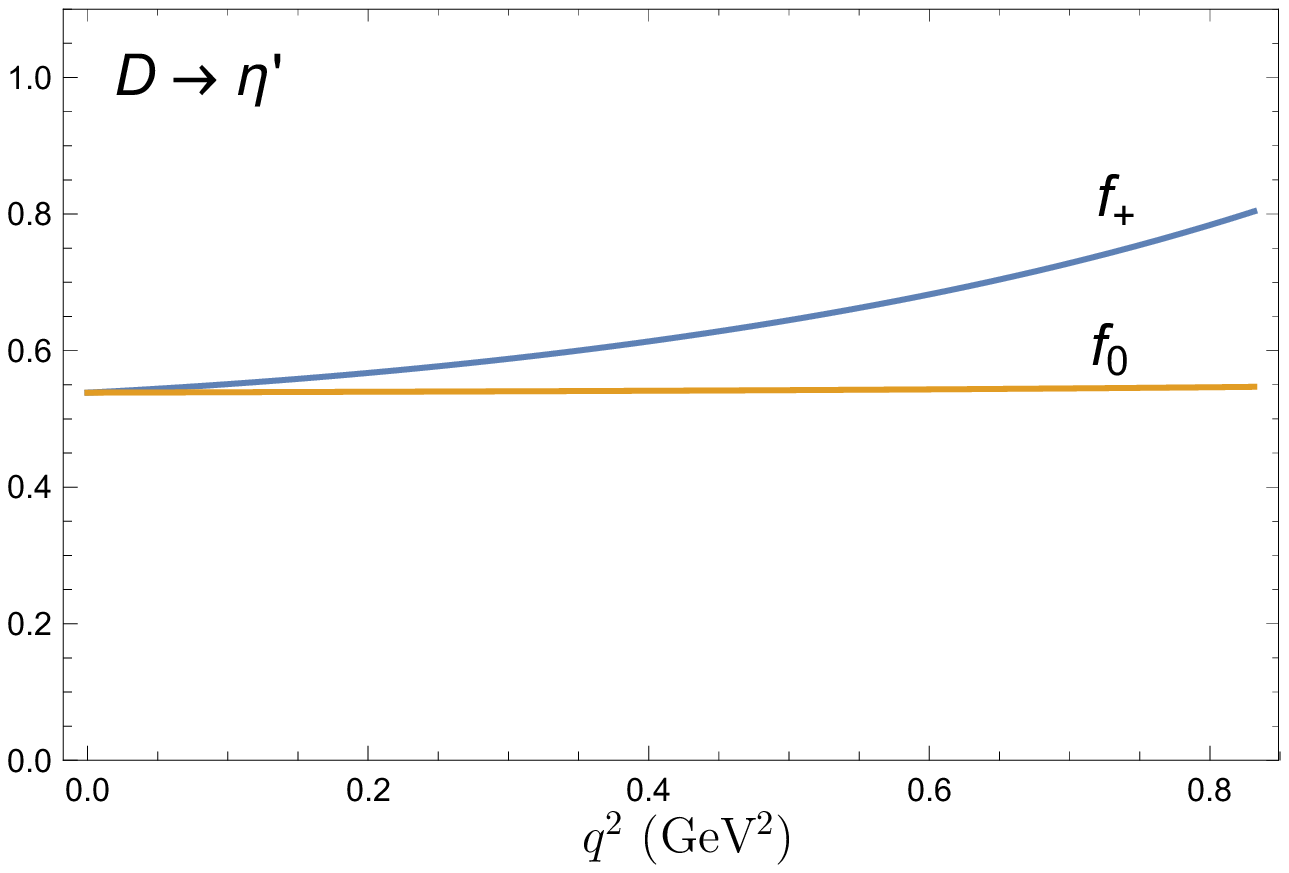}\\
  \includegraphics[width=8cm]{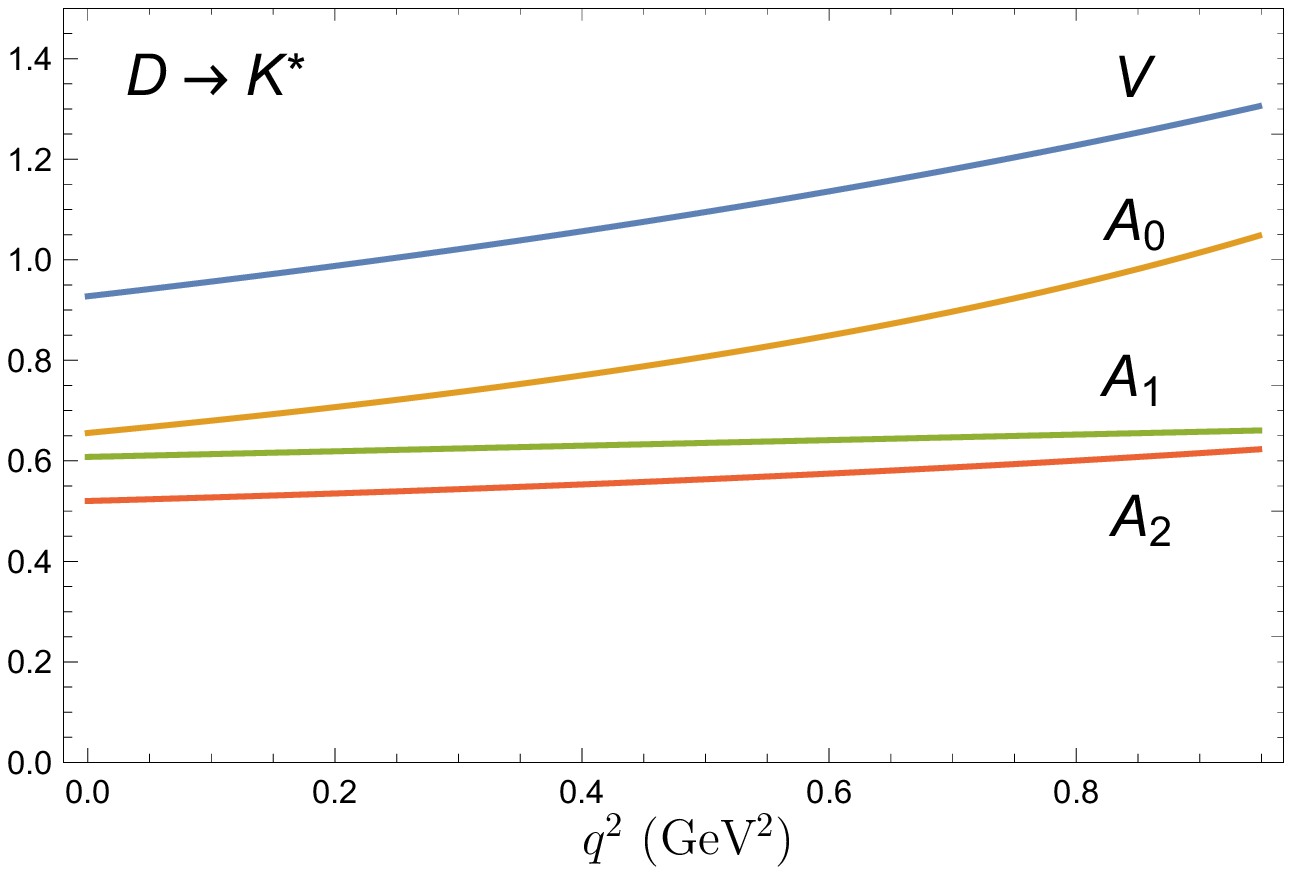}\ \
  \ \includegraphics[width=8cm]{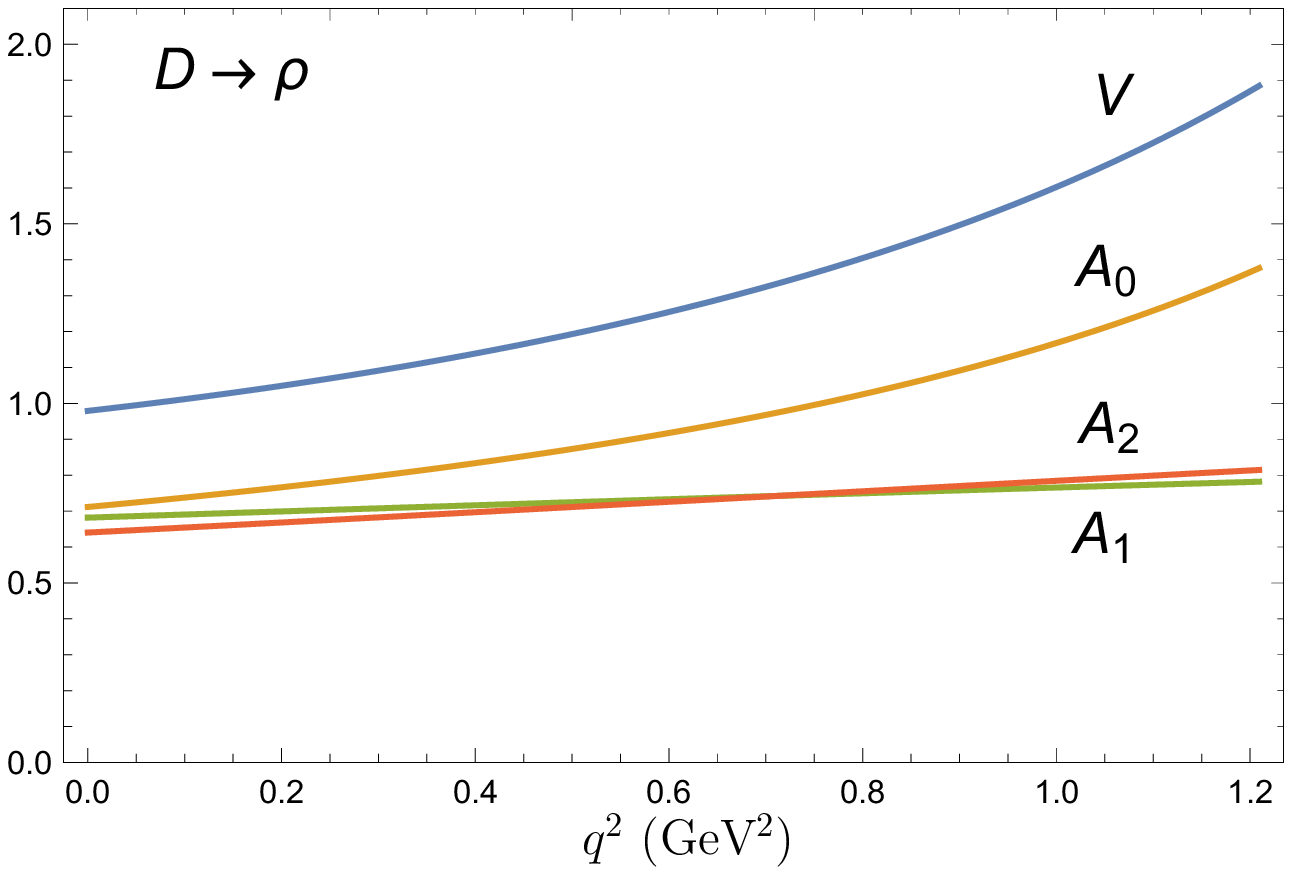}\\
  \includegraphics[width=8cm]{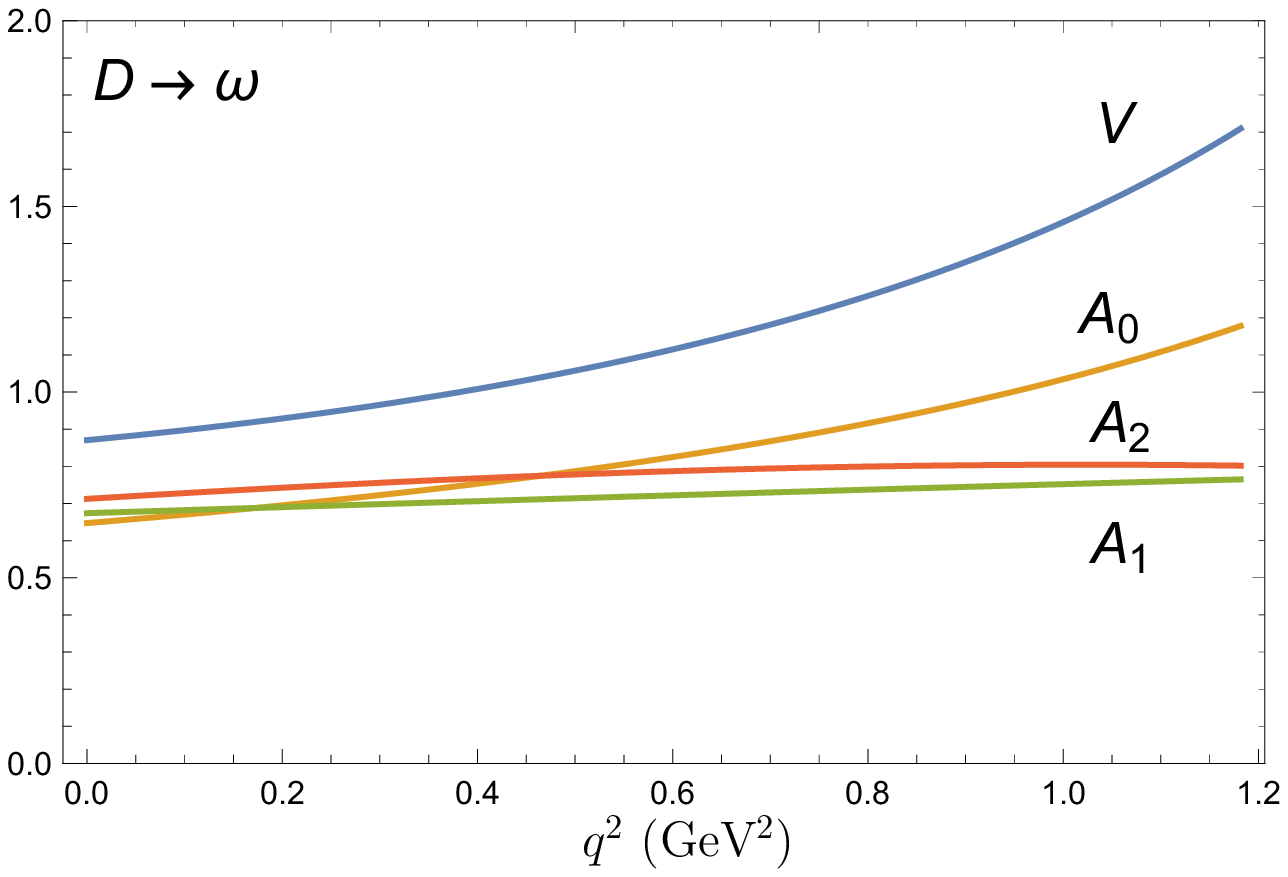}
\caption{Form factors of the weak $D$ meson transitions.    }
\label{fig:ffD}
\end{figure}

\begin{figure}
\centering
  \includegraphics[width=8cm]{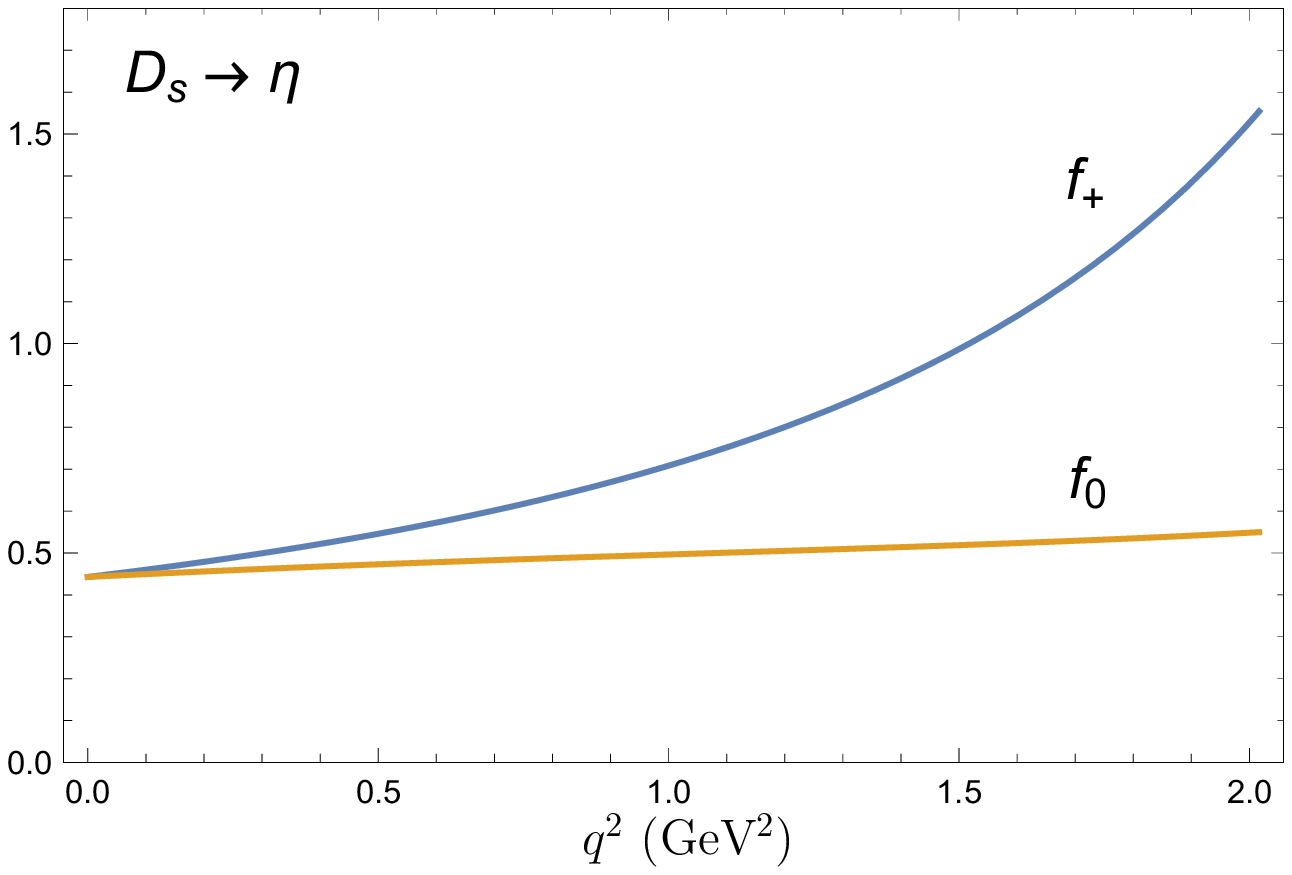}\ \
  \ \includegraphics[width=8cm]{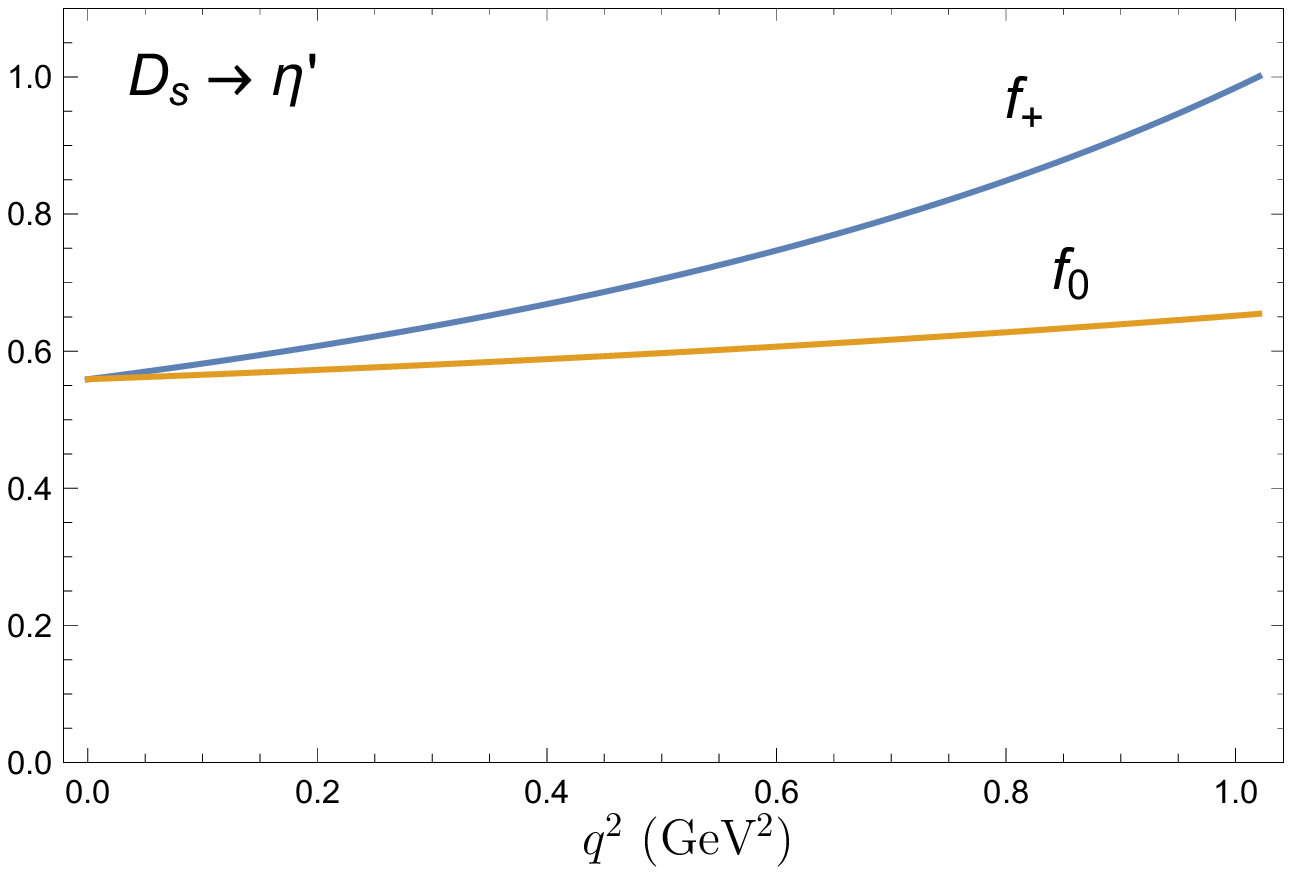}\\
  \includegraphics[width=8cm]{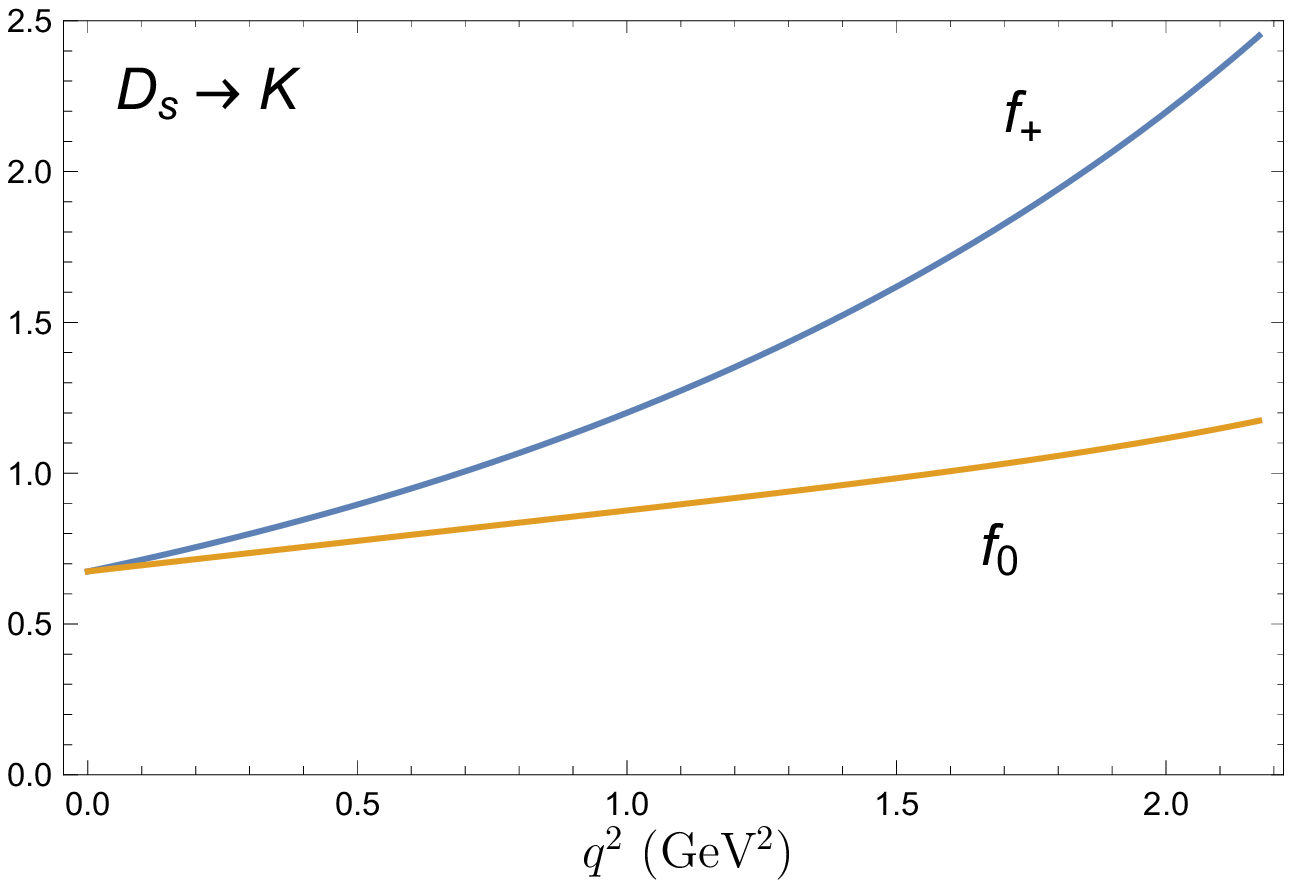}\ \
  \ \includegraphics[width=8cm]{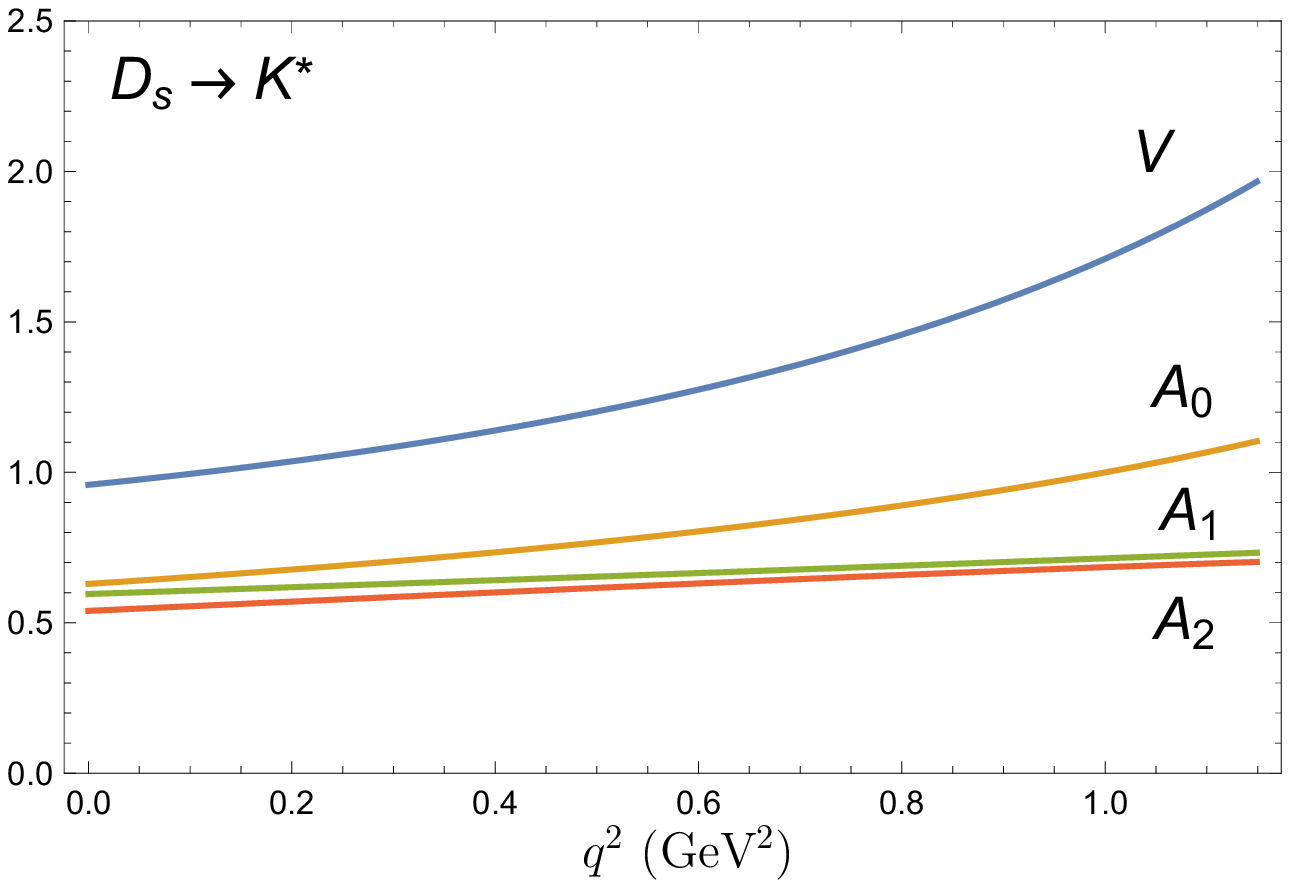}\\
  \includegraphics[width=8cm]{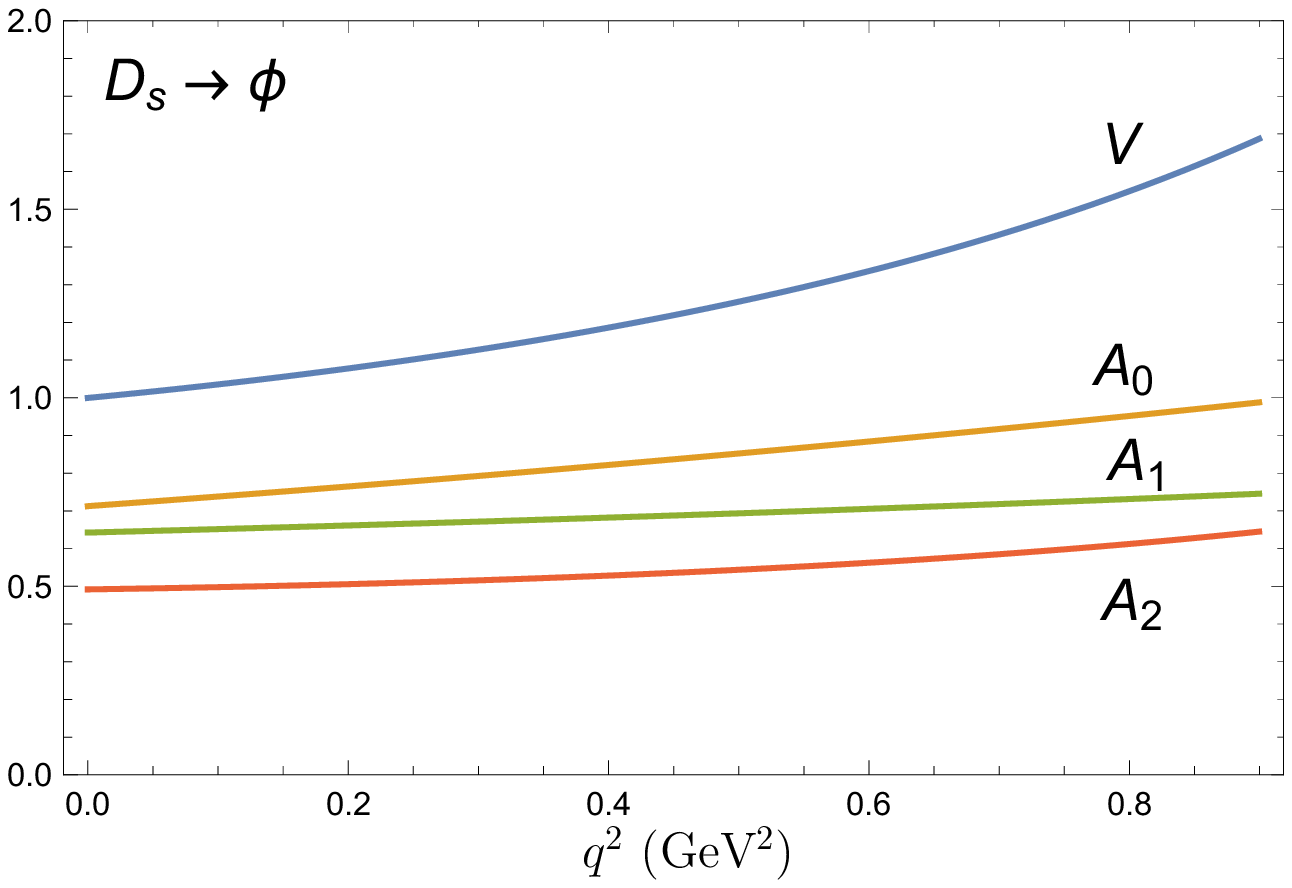}\ \
  \
\caption{Form factors of the weak $D_s$ meson transitions.    }
\label{fig:ffDs}
\end{figure}

(b) $f_0(q^2), A_1(q^2),A_2(q^2)$
\begin{equation}
  \label{fita12}
  F(q^2)=\frac{F(0)}{\displaystyle \left(1-\sigma_1
      \frac{q^2}{M_{D_{(s)}^*}^2}+ \sigma_2\frac{q^4}{M_{D_{(s)}^*}^4}\right)},
\end{equation}
where for the decays governed by the CKM favored $c\to s$ transitions
masses of the intermediate $D_s$ mesons are used:  $M=M_{D_{s}^*}=2.112$~GeV  for the form factors $f_+(q^2),V(q^2)$ and
$M=M_{D_s}=1.968$~GeV for the form factor $A_0(q^2)$. While for the
decays governed by the CKM
suppressed ($c\to d$) transitions masses of the intermediate $D$
mesons are taken as follows:  $M=M_{D^*}=2.010$~GeV  for the form factors $f_+(q^2),V(q^2)$ and
$M=M_{D}=1.870$~GeV for the form factor $A_0(q^2)$. The values of form
factors $F(0)$, $F(q^2_{\rm max})$ and fitted parameters
$\sigma_{1,2}$ are given in Tables~\ref{ffD},\ref{ffDs}. We estimate
the uncertainties of the calculated form factors to be less than 5\%. The
form factors are plotted in Figs.~\ref{fig:ffD},\ref{fig:ffDs}.

\begin{table}
\caption{Form factors of the weak $D$ meson transitions. }
\label{ffD}
\begin{ruledtabular}
\begin{tabular}{cccccc}
Decay& Form factor & $F(0)$& $F(q^2_{\rm max})$& $\sigma_1$ & $\sigma_2$ \\
\hline
$D\to K$         &$f_+$ &$0.716$ & $1.538$ & 0.902 & $1.07$\\
   & $f_0$  &$0.716$ & $1.086$ & 0.360& $1.657$ \\ \hline
$D\to K^*$   &$V$&$0.927$& $1.305$& $0.356$ &$-0.490$\\
     &$A_0$&$0.655$&$1.048$&$0.432$&$-0.840$\\
     &$A_1$&$0.608$& $0.660$& $0.410$ &$0.166$\\
     &$A_2$&$0.520$& $0.623$& $0.582$ &$-0.917$\\ \hline
$D\to \pi$   &$f_+$ &$0.640$ & $2.336$ & 0.332 & $0.557$\\
   & $f_0$  &$0.640$ & $1.318$ & $-0.345$& $1.133$ \\ \hline
$D\to \rho$   &$V$&$0.979$& $1.884$& $0.264$ &$-2.001$\\
     &$A_0$&$0.712$&$1.377$&$0.282$&$-0.826$\\
     &$A_1$&$0.682$& $0.782$& $0.567$ &$0.352$\\
     &$A_2$&$0.640$& $0.815$& $0.964$ &$0.645$\\\hline
$D\to \eta$   &$f_+$ &$0.547$ & $1.228$ & 1.153 & $1.519$\\
     & $f_0$  &$0.547$ & $0.683$ & $0.408$& $3.147$ \\ \hline
$D\to \eta'$   &$f_+$ &$0.538$ & $0.804$ & $-0.203$ & $-4.686$\\
   & $f_0$  &$0.538$ & $0.547$ &  $-0.950$& $1.038$ \\ \hline
$D\to \omega$   &$V$&$0.871$& $1.709$& $0.146$ &$-2.775$\\
     &$A_0$&$0.647$&$1.178$&$0.224$&$-0.759$\\
     &$A_1$&$0.674$& $0.765$& $0.542$ &$0.350$\\
     &$A_2$&$0.713$& $0.802$& $0.997$ &$2.176$\\
\end{tabular}
\end{ruledtabular}
\end{table}

\begin{table}
\caption{Form factors of the weak $D_s$ meson transitions. }
\label{ffDs}
\begin{ruledtabular}
\begin{tabular}{cccccc}
Decay& Form factor & $F(0)$& $F(q^2_{\rm max})$& $\sigma_1$ & $\sigma_2$ \\
\hline
$D_s\to \eta$   &$f_+$ &$0.443$ & $1.554$ & 0.675 & $-0.856$\\
     & $f_0$  &$0.443$ & $0.550$ & $-0.302$& $1.634$ \\ \hline
$D_s\to \eta'$   &$f_+$ &$0.559$ & $1.001$ & 0.719 & $-2.123$\\
   & $f_0$  &$0.559$ & $0.654$ &  $-0.499$& $-0.124$ \\ \hline
$D_s\to \phi$   &$V$&$0.999$& $1.687$& $0.467$ &$-4.020$\\
     &$A_0$&$0.713$&$0.988$&$0.412$&$0.903$\\
     &$A_1$&$0.643$& $0.746$& $0.621$ &$-0.317$\\
     &$A_2$&$0.492$& $0.645$& $0.447$ &$-3.622$\\ \hline
$D_s\to K$   &$f_+$ &$0.674$ & $2.451$ & 1.255 & $-0.935$\\
   & $f_0$  &$0.674$ & $1.174$ & $0.216$& $1.241$ \\ \hline
$D\to K^*$   &$V$&$0.959$& $1.966$& $0.425$ &$-2.444$\\
     &$A_0$&$0.629$&$1.103$&$0.281$&$-0.435$\\
     &$A_1$&$0.596$& $0.733$& $0.835$ &$0.423$\\
     &$A_2$&$0.540$& $0.702$& $1.266$ &$1.425$\\
\end{tabular}
\end{ruledtabular}
\end{table}

In Fig.~\ref{ffcomp} we compare our predictions for the product
$f_+(q^2)|V_{cq}|$  with experimental data
from Belle \cite{belleff}  and BaBar \cite{babarff}  and lattice results
 \cite{latt,latt-f} for the weak $D\to K$ and $D\to\pi$ transitions. On the
 same plots we also show our results for $f_0(q^2)|V_{cq}|$ in
 comparison with lattice  \cite{latt} data. We find the agreement
 within error bars with experimental values in the whole kinematical
 range for both transitions. There is also a nice accord with lattice
 results for the form factors of the $D\to \pi$ transition (there is a
 small difference only near $q^2_{\rm max}$), while for
 the $D\to K$ transition our form factors have systematically somewhat
 larger values for $q^2>0.7$~GeV$^2$. Note that in general our form
 factors better agree with data in the accessible kinematical range than lattice ones.

\begin{figure}
\centering
  \includegraphics[width=8cm]{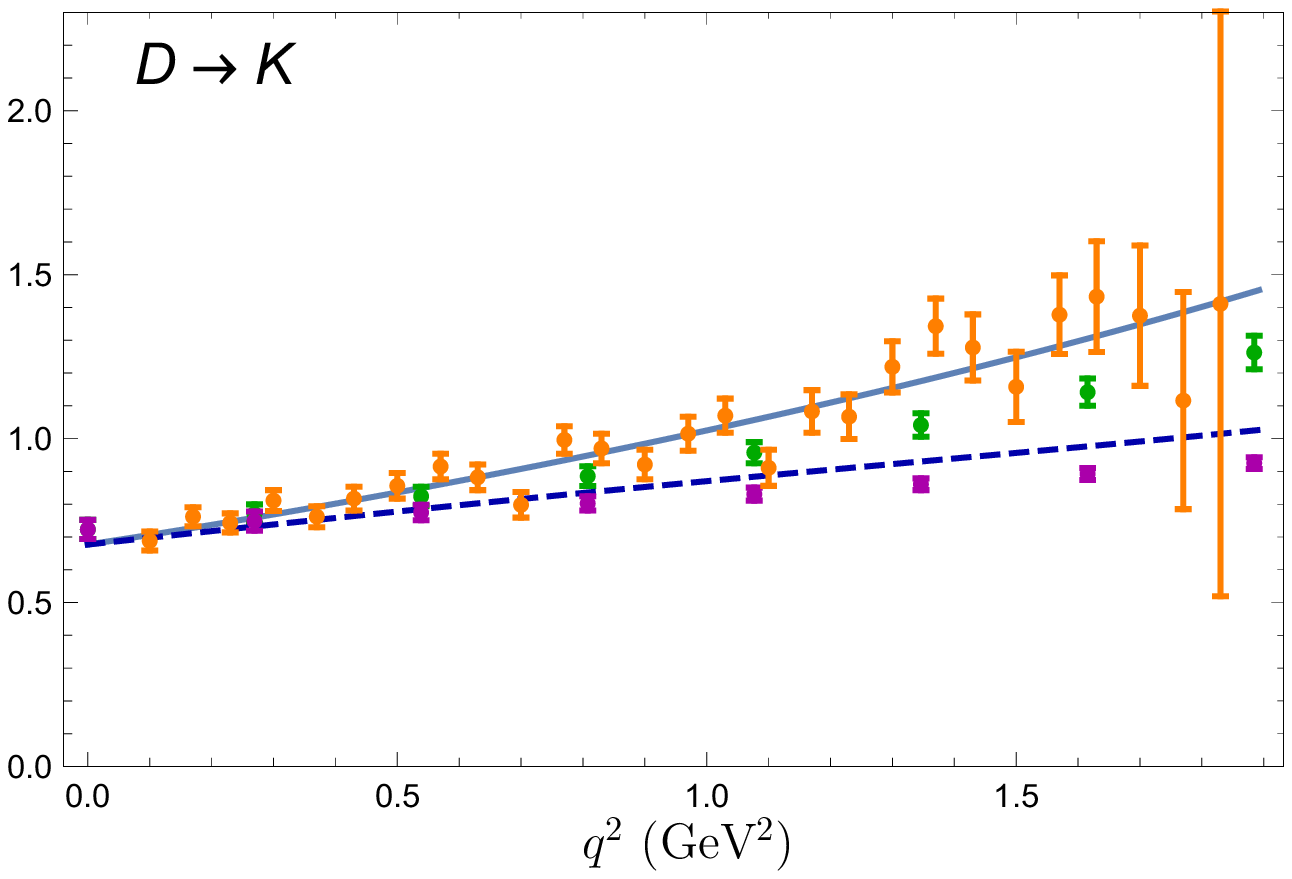}\ \
  \ \includegraphics[width=8cm]{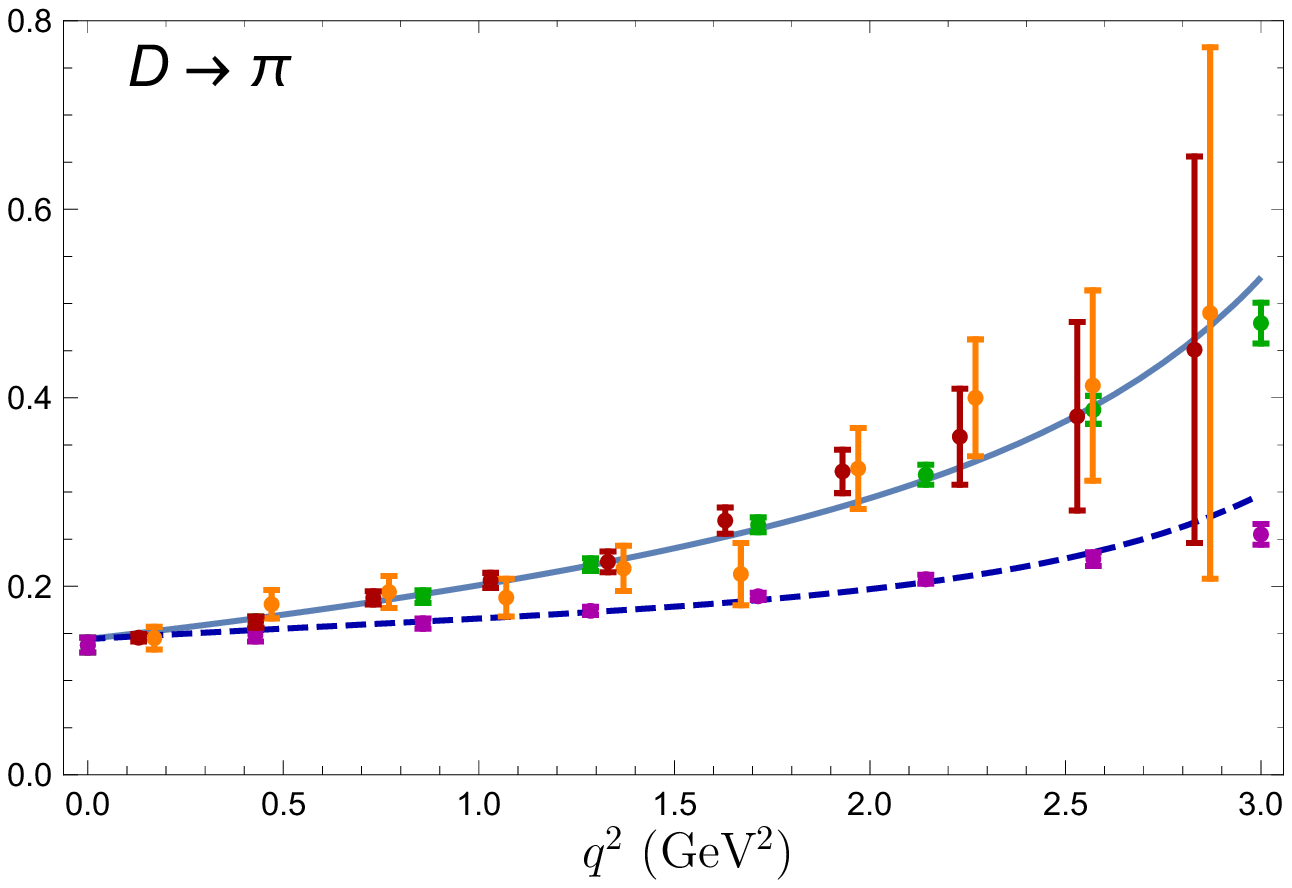}
\caption{Comparison of our predictions for the product  $f_+(q^2)|V_{cq}|$ (solid
  curve) and $f_0(q^2)|V_{cq}|$ (dashed curve)  with experimental data for $f_+(q^2)|V_{cq}|$
  form Belle \cite{belleff} (orange dots with error bars) and BaBar
  \cite{babarff} (red dots with error bars) and lattice results
  \cite{latt} for $f_+(q^2)|V_{cq}|$ (green dots with error bars) and $f_0(q^2)|V_{cq}|$
  (magenta dots with error bars) of the weak $D \to K$ ($q=s$) and
  $D\to\pi$ ($q=d$) transitions.    }
\label{fig:ffDcomp}
\end{figure}

In Table~\ref{ffcomp} we compare theoretical predictions for the form
factors of the weak $D$ and $D_s$ meson transitions  to pseudoscalar
mesons at $q^2=0$ with available experimental data. The authors of
Ref.~\cite{ikpsst} calculated form factors in the framework of the
covariant confining quark model. The covariant light-front quark model
was employed in Refs.~\cite{ck,verma,cchua}, while in Ref.~\cite{wzz} calculations
were done using light-cone sum rules in the framework of heavy quark
effective field theory. Lattice QCD simulations in Ref.~\cite{latt}
were carried out with $N_f=2+1+1$ dynamical quarks. Experimental
values were taken from very recent report on world averages of
measurements of hadron properties obtained by the Heavy Flavor
Averaging Group \cite{hfag}. Good agreement of our predictions with
data is found. Only for the $D\to\eta$ transition $f_+(0)$ is somewhat
larger than experimental value.

\begin{table}
\caption{Comparison of various theoretical predictions for the form factors
  $f_+(0)$ of the weak $D$ and $D_s$ meson transitions to pseudoscalar
  mesons with available
  experimental data. }
\label{ffcomp}
\begin{ruledtabular}
\begin{tabular}{ccccccc}
Decay& Our & \cite{ikpsst}& \cite{ck,verma,cchua}& \cite{wzz} &Lattice
\cite{latt}& Epxeriment \cite{hfag,besiii-6,besiii-7} \\
\hline
$D\to K$ &$0.716$ & $0.77$ & $0.79(1)$ & $0.661^{(67)}_{(66)}$&
 0.765(31)& $0.7361(34)$\\
$D\to \pi$ &$0.640$ & $0.63$ & 0.66(1) & $0.635^{(60)}_{(57)}$&
 0.612(35)& 0.6351(81)\\
$D\to \eta$  &$0.547$ & $0.36$ &0.55(1) &$0.556^{(56)}_{(53)}$ & &$0.38(3)$\\
$D\to \eta'$  &$0.538$ & $0.36$ &0.45(1)  & \\
$D_s\to \eta$   &$0.443$ & $0.49$ &0.48(3)  & $0.611^{(62)}_{(54)}$& &0.4576(70)\\
$D_s\to \eta'$  &$0.559$ & $0.59$ &0.59(3)  & &&0.490(51)\\
$D_s\to K$   &$0.674$ & $0.60$ & 0.66 & $0.820^{(80)}_{(71)}$& &0.720(85)\\
\end{tabular}
\end{ruledtabular}
\end{table}

For the weak $D^+$ and $D_s$ meson transitions to
vector mesons only the ratios of the form factors at maximum recoil of
the final meson ($q^2=0$) are obtained experimentally
\begin{equation}
  \label{eq:vfr}
  r_V=\frac{V(0)}{A_1(0)}, \qquad r_2=\frac{A_2(0)}{A_1(0)}.
\end{equation}
There have been many measurements and calculations of these ratios.
In Table~\ref{rcomp} we compare some of theoretical predictions with averaged
experimental data from PDG \cite{pdg} and recent BES III
\cite{besiii-1,besiii-2,besiii-3,besiii-4,besiii-5,besiii-6,besiii-7,besiii-8,besiii-9} measurements. Once again our results agree well with data.

\begin{table}
\caption{Comparison of various theoretical predictions for the ratios of the form factors
  $r_V=V(0)/A_1(0)$ and $r_2=A_2(0)/A_1(0)$  of the weak $D^+$ and $D_s$ meson transitions to vector mesons with available
  experimental data. }
\label{rcomp}
\begin{ruledtabular}
\begin{tabular}{cccccccc}
Decay& Ratio& \multicolumn{4}{c}{Theory} & \multicolumn{2}{c}{
                Epxeriment} \\\cline{3-6} \cline{7-8}
  && Our & \cite{ikpsst}& \cite{ck,verma,cchua}& \cite{wzz}& PDG \cite{pdg}&BES
  III \cite{besiii-1,besiii-2,besiii-3,besiii-4,besiii-5,besiii-6,besiii-7,besiii-8,besiii-9}\\
\hline
$D\to K^*$ &$r_V$&$1.53$ & $1.22(24)$ & $1.36(2)$ & $1.39^{(9)}_{(10)}$&
                                                                         1.49(5)& $1.406(62)$\\
  &$r_2$&$0.85$ & $0.92(18)$ & $0.83(3)$ & $0.60^{(9)}_{(8)}$&
 0.802(21)& $0.784(48)$\\ \hline
$D\to \rho$ &$r_V$&$1.44$ & $1.26(25)$ & 1.46(3) & $1.34^{(14)}_{(13)}$&
                                                                         1.48(16)& 1.695(98)\\
   &$r_2$&$0.94$ & $0.93(19)$ & $0.78(2)$ & $0.62^{(8)}_{(8)}$&
                                                                 0.83(12)& $0.845(68)$\\ \hline
  $D\to \omega$ &$r_V$&$1.29$ & $1.24(25)$ & 1.47(4) & $1.33^{(15)}_{(13)}$
 & 1.24(11)\\
   &$r_2$&$1.05$ & $0.95(19)$ & $0.84(2)$ & $0.60^{(9)}_{(9)}$
                                   & $1.06(16)$\\ \hline
$D_s\to \phi$ &$r_V$&$1.56$ & $1.34(27)$ & $1.42(2)$ & $1.37^{(24)}_{(21)}$&
                                                                         1.80(8)& \\
  &$r_2$&$0.77$ & $0.99(20)$ & $0.86(1)$ & $0.53^{(10)}_{(6)}$&
                                                                0.84(11)& \\ \hline
  $D_s\to K^*$ &$r_V$&$1.61$ & $1.40(28)$ & $1.55(5)$ & $1.31^{(19)}_{(16)}$&
                                                                         &1.67(38)\\
  &$r_2$&$0.90$ & $0.99(20)$ & $0.82(2)$ & $0.53^{(10)}_{(6)}$&&
 0.77(29)\\
\end{tabular}
\end{ruledtabular}
\end{table}

\section{Semileptonic decays}
\label{sec:sd}

The differential decay rate of the semileptonic $D_{(s)}$ decays can
be expressed in the following form \cite{ikpsst}
\begin{eqnarray}
  \label{eq:dGamma}
  \frac{d\Gamma(D_{(s)}\to
  F\ell^+\nu_\ell)}{dq^2d(\cos\theta)}&=&\frac{G_F^2}{(2\pi)^3}
  |V_{cq}|^2\frac{\lambda^{1/2}(q^2-m_\ell^2)^2}{64M_{D_{(s)}}^3q^2}\Biggl[(1+\cos^2\theta)
{\cal H}_U+2\sin^2\theta{\cal H}_L+2\cos\theta{\cal H}_P\cr
&&+\frac{m_\ell^2}{q^2}(\sin^2\theta{\cal H}_U
   +2\cos^2\theta{\cal H}_L+ 2 {\cal H}_S-4\cos\theta {\cal H}_{SL})\Bigr],
\end{eqnarray}
where $\lambda\equiv
\lambda(M_{D_{(s)}}^2,M_F^2,q^2)=M_{D_{(s)}}^4+M_F^4+q^4-2(M_{D_{(s)}}^2M_F^2+M_F^2q^2+M_{D_{(s)}}^2q^2)$,
$m_\ell$ is the lepton mass, and the polar angle $\theta$ is the angle
between the momentum of the charged lepton in the rest frame of the intermediate
$W$-boson and the direction opposite
to the final $F$ meson momentum in the rest frame of $D_{(s)}$.  The bilinear combinations  ${\cal H}_I$ of the helicity components of the
hadronic tensor are defined by \cite{ikpsst}
\begin{equation}
  \label{eq:hh}
{\cal H}_U=|H_+|^2+|H_-|^2, \quad {\cal H}_L=|H_0|^2, \quad {\cal
   H}_P=|H_+|^2-|H_-|^2,\quad
    {\cal H}_S=|H_t|^2, \quad {\cal H}_{SL}=\Re(H_0H_t^\dag),
  \end{equation}
and the helicity amplitudes are expressed through invariant form
factors.
\begin{itemize}
\item For $D_{(s)}\to P$ transitions
  \begin{equation}
  \label{eq:hap}
  H_\pm=0,\quad
H_0=\frac{\lambda^{1/2}}{\sqrt{q^2}}f_+(q^2),\quad
H_t=\frac1{\sqrt{q^2}}(M_{D_{(s)}}^2-M_{P}^2)f_0(q^2).
\end{equation}
\item For $D_{(s)}\to V$ transitions
  \begin{eqnarray}
  \label{eq:hav}
\!\!\!\!\!\!\!\!\!\!\!\!\!\!\!  H_\pm(q^2)&=&\frac{\lambda^{1/2}}{M_{D_{(s)}}+M_{V}}\left[V(q^2)\mp
\frac{(M_{D_{(s)}}+M_{V})^2}{\lambda^{1/2}}A_1(q^2)\right],\cr
\!\!\!\!\!\!\!\!\!\!\!\!\!\!\!    H_0(q^2)&=&\frac1{2M_{V}\sqrt{q^2}}\left[(M_{D_{(s)}}+M_{V})
(M_{D_{(s)}}^2-M_{V}^2-q^2)A_1(q^2)-\frac{\lambda}{M_{D_{(s)}}
+M_{V}}A_2(q^2)\right], \cr
\!\!\!\!\!\!\!\!\!\!\!\!\!\!\!    H_t&=&\frac{\lambda^{1/2}}{\sqrt{q^2}}A_0(q^2).
\end{eqnarray}
\end{itemize}

The expression (\ref{eq:dGamma}) normalized by the  decay rate
$d\Gamma/dq^2$, which is obtained by the integration of (\ref{eq:dGamma}) over $\cos\theta$,
can be rewritten as
\begin{equation}
  \label{eq:ndr}
\frac1{d\Gamma/dq^2} \frac{d\Gamma(D_{(s)}\to
  F\ell^+\nu_\ell)}{dq^2d(\cos\theta)} = \frac12\left[1-\frac13
  C^\ell_F(q^2)\right] +A_{FB}(q^2)\cos\theta+\frac12 C^\ell_F(q^2)\cos^2\theta,
\end{equation}
where the forward-backward asymmetry is defined by
\begin{equation}
  \label{eq:afb}
  A_{FB}(q^2)=\frac{\int^1_0d(\cos\theta)d\Gamma/d(\cos\theta)-\int^0_{-1}d(\cos\theta)d\Gamma/d(\cos\theta)}{\int^1_0d(\cos\theta)d\Gamma/d(\cos\theta)+\int^0_{-1}d(\cos\theta)d\Gamma/d(\cos\theta)}=\frac34\frac{{\cal
      H}_P-2\frac{m_\ell^2}{q^2}{\cal H}_{SL}}{{\cal H}_{\rm total}},
\end{equation}
and lepton-side convexity parameter, which is the second derivative of
the distribution (\ref{eq:ndr}) over $\cos\theta$, is given by
\begin{equation}
  \label{eq:clf}
   C^\ell_F(q^2)=\frac34\left(1-\frac{m_\ell^2}{q^2}\right)\frac{{\cal H}_U-2{\cal H}_L}{{\cal H}_{\rm total}}.
\end{equation}
Here the total helicity structure
\begin{equation}
  \label{eq:htot}
 {\cal H}_{\rm total}= ({\cal H}_U+{\cal
   H}_L)\left(1+\frac{m_\ell^2}{2q^2}\right) +\frac{3m_\ell^2}{2q^2}{\cal H}_S
\end{equation}
enters the differential decay distribution (\ref{eq:dGamma}) integrated
over   $\cos\theta$
\begin{equation}
  \label{eq:dg}
   \frac{d\Gamma(D_{(s)}\to
  F\ell^+\nu_\ell)}{dq^2}=\frac{G_F^2}{(2\pi)^3}
  |V_{cq}|^2\frac{\lambda^{1/2}(q^2-m_\ell^2)^2}{24M_{D_{(s)}}^3q^2}{\cal H}_{\rm total}.
\end{equation}

Other useful observables are the longitudinal polarization of the final
charged lepton $\ell$ defined by  \cite{ikpsst}
\begin{equation}
  \label{eq:ple}
  P_L^\ell(q^2)=\frac{({\cal H}_U+{\cal
   H}_L)\left(1-\frac{m_\ell^2}{2q^2}\right) -\frac{3m_\ell^2}{2q^2}{\cal H}_S}{{\cal H}_{\rm total}},
\end{equation}
and its transverse polarization  \cite{ikpsst}
\begin{equation}
  \label{eq:pte}
  P_T^\ell(q^2)=-\frac{3\pi m_\ell}{8\sqrt{q^2}}\frac{{\cal H}_P+2{\cal H}_{SL}}{{\cal H}_{\rm total}}.
\end{equation}
For the decays $D_{(s)}\to V$ the longitudinal polarization fraction
of the final vector meson is given by \cite{ikpsst}
\begin{equation}
  \label{eq:fl}
  F_L(q^2)=\frac{{\cal
   H}_L\left(1+\frac{m_\ell^2}{2q^2}\right) +\frac{3m_\ell^2}{2q^2}{\cal H}_S}{{\cal H}_{\rm total}},
\end{equation}
then its transverse polarization fraction $F_T(q^2)=1- F_L(q^2)$.

\section{Results and discussion}
\label{sec:rd}

\begin{figure}
\centering
  \includegraphics[width=8cm]{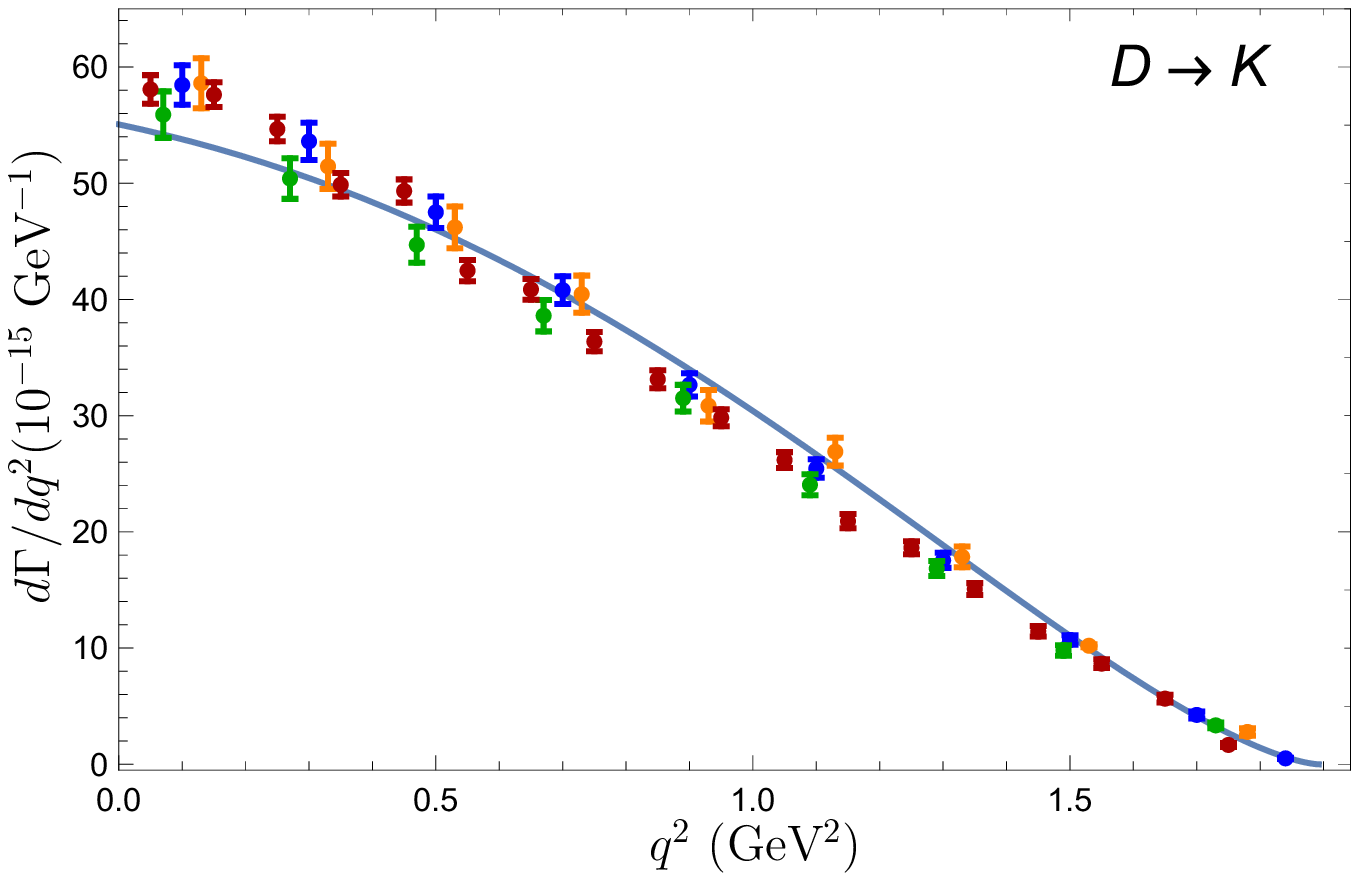}\ \
  \ \includegraphics[width=8cm]{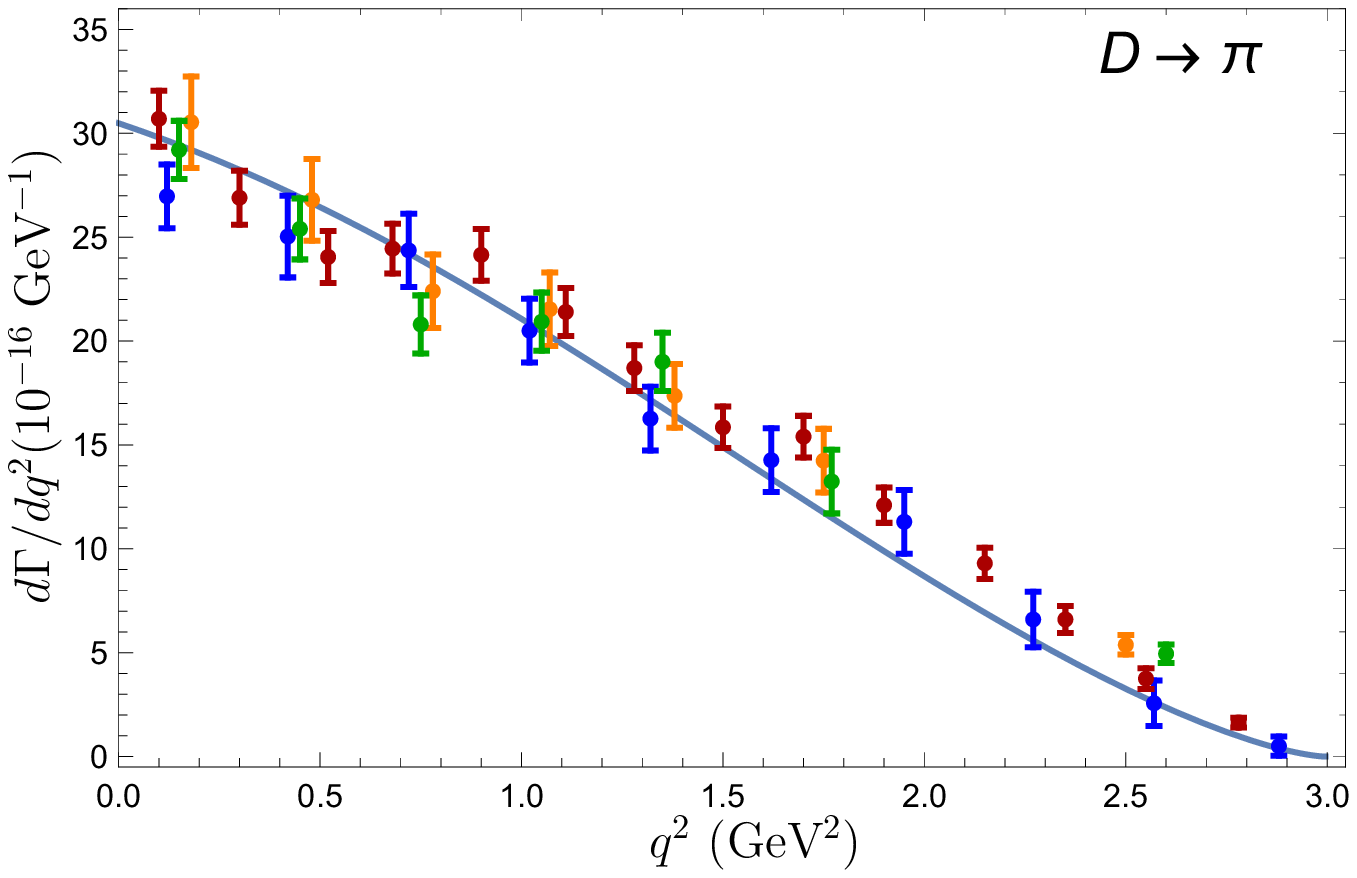}
\caption{Comparison of our predictions for the weak $D \to K e\nu_e$  and
  $D\to\pi e\nu_e$ differential decay
  rates   with experimental data
  form  BaBar
  \cite{babarff,babarff2} (blue dots with error bars), CLEO \cite{cleo} (orange
  dots with error bars) and BES III
  \cite{besiii-1,besiiio-1} for   neutral $D^0$ (red dots with error bars) and charged $D^+$ with the account of isospin factor (green dots with error bars) .   }
\label{fig:GDcomp}
\end{figure}

Now we substitute the form factors calculated in  Sec.~\ref{sec:dff}
in the expressions for helicity amplitudes, Eq.~(\ref{eq:hap}) and
Eq.~(\ref{eq:hav}), and then evaluate differential and total decay rates of
semileptonic $D_{(s)}$ decays. In Fig.~\ref{fig:GDcomp} we confront our
results with experimental data from  BaBar \cite{babarff},  CLEO
\cite{cleo} and BESIII \cite{besiii-1,besiiio-1} Collaborations for $D \to K e\nu_e$  and $D\to\pi e\nu_e$
differential decay  rates. Good agreement in the whole accessible
kinematical range is observed. In
Tables~\ref{brDKcomp}-\ref{brDsKcomp} we compare our and previous
theoretical predictions \cite{ikpsst,ck,wzz} with experimental data
from PDG \cite{pdg} and recent data from BES III \cite{besiii-1,besiii-2,besiii-3,besiii-4,besiii-5,besiii-6,besiii-7,besiii-8,besiii-9}
Collaboration. We roughly estimate the uncertainties of our
calculations to be within 10\%. For all decays we find agreement with
experimental data within error bars. Only for the $D\to K^*\ell\nu_\ell$ decay branching fractions we obtain somewhat lower central values  than the data \cite{pdg}, while Refs.~\cite{ikpsst,ck} give significantly larger values.  Thus the precise measurement of these branching fractions is the important test of the models.

\begin{table}
\caption{Comparison of  various theoretical predictions for the branching
  ratios (in \%)  of the CKM-favoured $D\to K^{(*)}\ell\nu_\ell$ semileptonic decays with available
  experimental data. }
\label{brDKcomp}
\begin{ruledtabular}
\begin{tabular}{ccccccc}
Decay& \multicolumn{4}{c}{Theory} & \multicolumn{2}{c}{
                Epxeriment} \\\cline{2-5} \cline{6-7}
  & Our & \cite{ikpsst}& \cite{ck}& \cite{wzz}& PDG \cite{pdg}&BES
  III \cite{besiii-1,besiii-5,besiii-8}\\
\hline
$D^+\to \bar K^0e^+\nu_e$ &$9.02$ & $9.28$ & $10.32(93)$ &
                                                           $8.12^{(1.19)}_{(1.08)}$&  8.73(10)& 8.60(16) \\
$D^+\to \bar K^0\mu^+\nu_\mu$  &$8.85$ & $9.02$ & $10.07(91)$ & $7.98^{(1.16)}_{(1.06)}$&
                                                                                         8.76(19)& \\
$D^0\to  K^-e^+\nu_e$ &$3.56$ & $3.63$ &4.1(4)   & $3.20^{(47)}_{(43)}$&
                                                                 3.542(35)&
  3.505(36)\\
$D^0\to K^-\mu^+\nu_\mu$  &$3.49$ & $3.53$ &4.2(4)  & $3.10^{(46)}_{(42)}$&
                                                                     3.41(4)& 3.413(39) \\
$D^+\to \bar K^{*0}e^+\nu_e$ &$4.87$ & $7.61$ & $7.5(7)$ & $5.37^{(24)}_{(23)}$&  5.40(10)& \\
$D^+\to \bar K^{*0}\mu^+\nu_\mu$  &$4.62$ & $7.21$ & $7.0(7)$ & $5.10^{(23)}_{(21)}$& 5.27(15)& \\
$D^0\to  K^{*-}e^+\nu_e$ &$1.92$ & $2.96$ &3.0(3)  & $2.12^{(9)}_{(9)}$&
                                                                 2.15(16)&
  2.033(66)\\
$D^0\to K^{*-}\mu^+\nu_\mu$  &$1.82$ & $2.80$ &2.8(3)  & $2.01^{(9)}_{(8)}$&
 1.89(24)&  \\
\end{tabular}
\end{ruledtabular}
\end{table}

\begin{table}
\caption{Comparison of various theoretical predictions for the branching
  ratios (in $10^{-3}$)  of the CKM-suppressed $D$ meson semileptonic decays with available
  experimental data. }
\label{brDpicomp}
\begin{ruledtabular}
\begin{tabular}{ccccccc}
Decay& \multicolumn{4}{c}{Theory} & \multicolumn{2}{c}{
                Epxeriment} \\\cline{2-5} \cline{6-7}
  & Our & \cite{ikpsst}& \cite{ck}& \cite{wzz}& PDG \cite{pdg}&BES
  III \cite{besiii-2,besiii-4,besiii-9}\\
\hline
$D^+\to \pi^0e^+\nu_e$ &$3.53$ & $2.9$ & $4.1(3)$ &
                                                    $3.52^{(45)}_{(38)}$&  3.72(17)& 3.63(9) \\
$D^+\to \pi^0\mu^+\nu_\mu$  &$3.47$ & $2.8$ & $4.1(3)$ & $3.49^{(45)}_{(38)}$&
  3.50(15)& 3.50(15)\\
$D^0\to  \pi^-e^+\nu_e$ &$2.78$ & $2.2$ &3.2(3)  & $2.78^{(35)}_{(30)}$&
 2.91(4)& 2.95(5)\\
$D^0\to \pi^-\mu^+\nu_\mu$  &$2.74$ & $2.2$ &3.2(3)  & $2.75^{(35)}_{(30)}$&            2.67(12)& 2.72(10) \\
$D^+\to \rho^0 e^+\nu_e$ &$2.49$ & $2.09$ & $2.3(2)$ & $2.29^{(23)}_{(16)}$&  $2.18^{(17)}_{(25)}$&1.860(93) \\
$D^+\to \rho^{0}\mu^+\nu_\mu$  &$2.39$ & $2.01$ & $2.2(2)$ & $2.20^{(21)}_{(16)}$& 2.4(4)& \\
$D^0\to  \rho^{-}e^+\nu_e$ &$1.96$ & $1.62$ &1.8(2)  & $1.81^{(18)}_{(13)}$&
 1.77(16)&  1.445(70)\\
$D^0\to \rho^-\mu^+\nu_\mu$  &$1.88$ & $1.55$ &1.7(2)  & $1.73^{(17)}_{(13)}$&
                                                                         1.89(24)&  \\
$D^+\to \eta e^+\nu_e$ &$1.24$ & $0.938$ & $1.2(1)$ & $0.86^{(16)}_{(15)}$&  1.11(7)& 1.074(98) \\
$D^+\to \eta\mu^+\nu_\mu$  &$1.21$ & $0.912$ & $1.2(1)$ & $0.84^{(16)}_{(14)}$&
                                  & \\
  $D^+\to \eta' e^+\nu_e$ &$0.225$ & $0.200$ & $0.18(2)$ & &  0.20(4)& 0.191(53) \\
$D^+\to \eta'\mu^+\nu_\mu$  &$0.211$ & $0.190$ & $0.17(2)$ & &  & \\
$D^+\to \omega e^+\nu_e$ &$2.17$ & $1.85$ & $2.1(2)$ & $1.93^{(20)}_{(14)}$&  $1.69(11)$ &2.05(72) \\
$D^+\to \omega\mu^+\nu_\mu$  &$2.08$ & $1.78$ & $2.0(2)$ & $1.85^{(19)}_{(13)}$& & \\
\end{tabular}
\end{ruledtabular}
\end{table}

\begin{table}
\caption{Comparison of  various theoretical predictions for the branching
  ratios (in \%)  of the $D_s$ meson semileptonic decays with available
  experimental data. }
\label{brDsKcomp}
\begin{ruledtabular}
\begin{tabular}{ccccccc}
Decay& \multicolumn{4}{c}{Theory} & \multicolumn{2}{c}{
                Epxeriment} \\\cline{2-5} \cline{6-7}
  & Our & \cite{ikpsst}& \cite{ck}& \cite{wzz}& PDG \cite{pdg}&BES
  III \cite{besiii-3,besiii-6,besiii-7}\\
\hline
$D_s\to \eta e^+\nu_e$ &$2.37$ & $2.24$ & $2.26(21)$ &
                                                    $1.27^{(26)}_{(20)}$&  2.29(19)& 2.323(89) \\
$D_s\to \eta\mu^+\nu_\mu$  &$2.32$ & $2.18$ & $2.22(20)$ & $1.25^{(25)}_{(20)}$&
  2.4(5)& 2.42(47)\\
$D_s\to \eta' e^+\nu_e$ &$0.87$ & $0.83$ & $0.89(9)$ &&  0.74(14)& 0.824(78) \\
$D_s\to \eta'\mu^+\nu_\mu$  &$0.83$ & $0.79$ & $0.85(8)$ & &
  1.1(5)& 1.06(54)\\
$D_s\to \phi e^+\nu_e$ &$2.69$ & $3.01$ & $3.1(3)$ & $2.53^{(37)}_{(40)}$&  2.39(16)& 2.26(46) \\
$D_s\to \eta\mu^+\nu_\mu$  &$2.54$ & $2.85$ & $2.9(3)$ & $2.40^{(35)}_{(37)}$&
  1.9(5)& 1.94(54)\\
$D_s\to K^0 e^+\nu_e$ &$0.40$ & $0.20$ & $0.27(2)$ &$0.390^{(74)}_{(57)}$&  0.39(9)& 0.325(4) \\
$D_s\to K^0\mu^+\nu_\mu$  &$0.39$ & $0.20$ & $0.26(2)$ &$0.383^{(72)}_{(56)}$ &
  & \\
$D_s\to K^{*0} e^+\nu_e$ &$0.21$ & $0.18$ & $0.19(2)$ &$0.233^{(29)}_{(30)}$&  0.18(4)& 0.237(33) \\
$D_s\to K^{*0}\mu^+\nu_\mu$  &$0.20$ & $0.17$ & $0.19(2)$ &$0.224^{(27)}_{(29)}$ &
  & \\
\end{tabular}
\end{ruledtabular}
\end{table}

Recently possible hints of the violation of the lepton universality
were found in $B$ decays where deviations from the standard model
predictions for the ratios of the semileptonic decay
branching fractions involving muon and electron were observed. In
Table~\ref{lut} we give our results for the corresponding ratios of
$D$ decays
\begin{equation}
  \label{eq:R}
  R_F=\frac{\Gamma(D_{(s)}\to F\mu^+\nu_\mu)}{\Gamma(D_{(s)}\to F e^+\nu_e)}
\end{equation}
in comparison with previous predictions \cite{ikpsst,ck}, lattice
\cite{latt} and experimental data form the BES III  Collaboration
\cite{besiii-2,besiii-3,besiii-5,hfag}. We see that the standard model predictions are
consistent with current experimental data.

\begin{table}
\caption{Test of the $e-\mu$ lepton flavor universality. Comparison of theoretical
  predictions for the ratios $R$ of  the weak $D$ and $D_s$ meson
  semileptonic decays 
  with available
  experimental data. }
\label{lut}
\begin{ruledtabular}
\begin{tabular}{cccccc}
Decay& Our & \cite{ikpsst}& \cite{ck}  &Lattice
\cite{latt-f}& Epxeriment \cite{besiii-2,besiii-3,besiii-5,hfag} \\
\hline
$D\to K$ &$0.980$ & $0.97$ & $0.976$ & $0.975(1)$& $\left\{
\begin{array}{ll}0.974(14)&K^-\\1.00(3)&\bar K^0\end{array}\right.$\\
$D\to \pi$ &$0.985$ & $0.98$ & 1.00 & 0.985(2)&$\left\{
\begin{array}{ll}
  0.964(40)&\pi^0\\
0.922(40)&\pi^-
\end{array}\right.$\\
  $D\to K^*$ &$0.946$ & $0.95$ & $0.933$ & & \\
 $D\to \rho$ &$0.959$ & $0.96$ & $0.957$ & & \\
$D\to \eta$  &$0.976$ & $0.97$&  1.00& & \\
  $D\to \eta'$  &$0.937$ & $0.95$ &0.944  & \\
  $D\to \omega$ &$0.959$ & $0.96$ & $0.952$ & & \\
$D_s\to \eta$   &$0.977$ & $0.97$ &0.982  &  &1.05(24)\\
  $D_s\to \eta'$  &$0.952$ & $0.95$ &0.956   &&1.14(68)\\
  $D_s\to \phi$  &$0.944$ & $0.95$ &0.936   &&0.86(29)\\
  $D_s\to K$   &$0.984$ & $1.00$ & 0.963 &  &\\
  $D_s\to K^*$   &$0.958$ & $0.95$ & 1.00 &  &\\
\end{tabular}
\end{ruledtabular}
\end{table}

\begin{figure}
\centering
  \includegraphics[width=8cm]{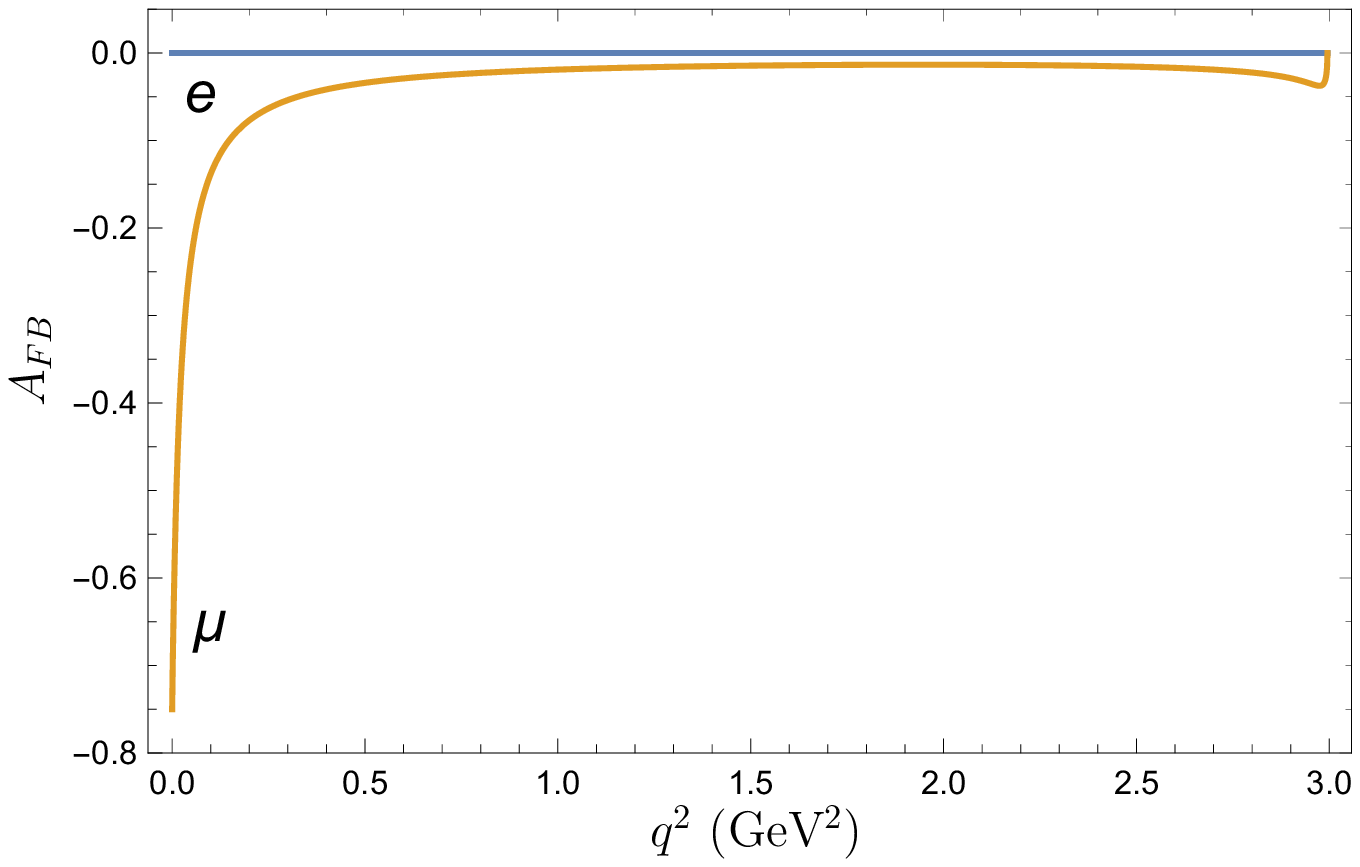}\ \
  \ \includegraphics[width=8cm]{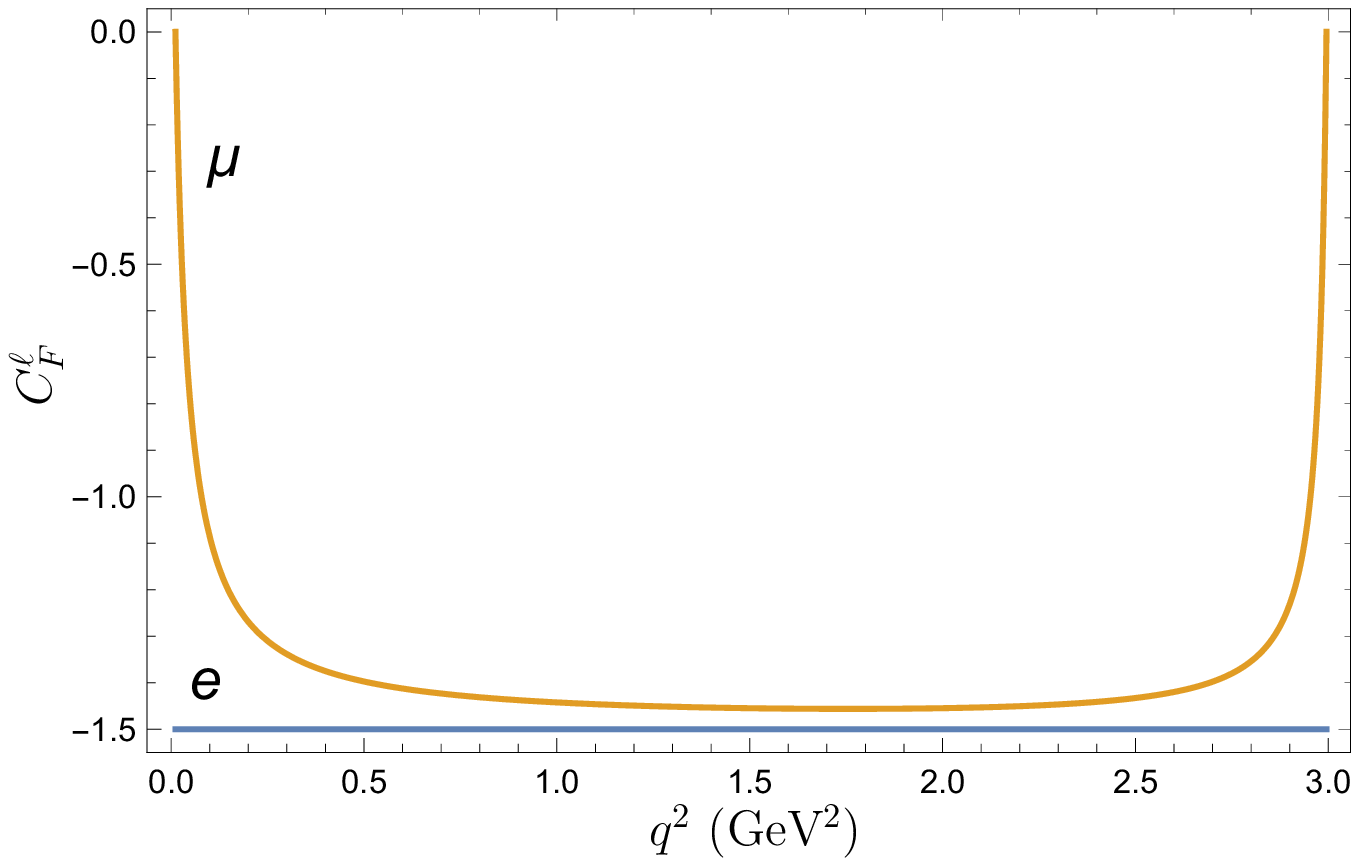}\\
  \includegraphics[width=8cm]{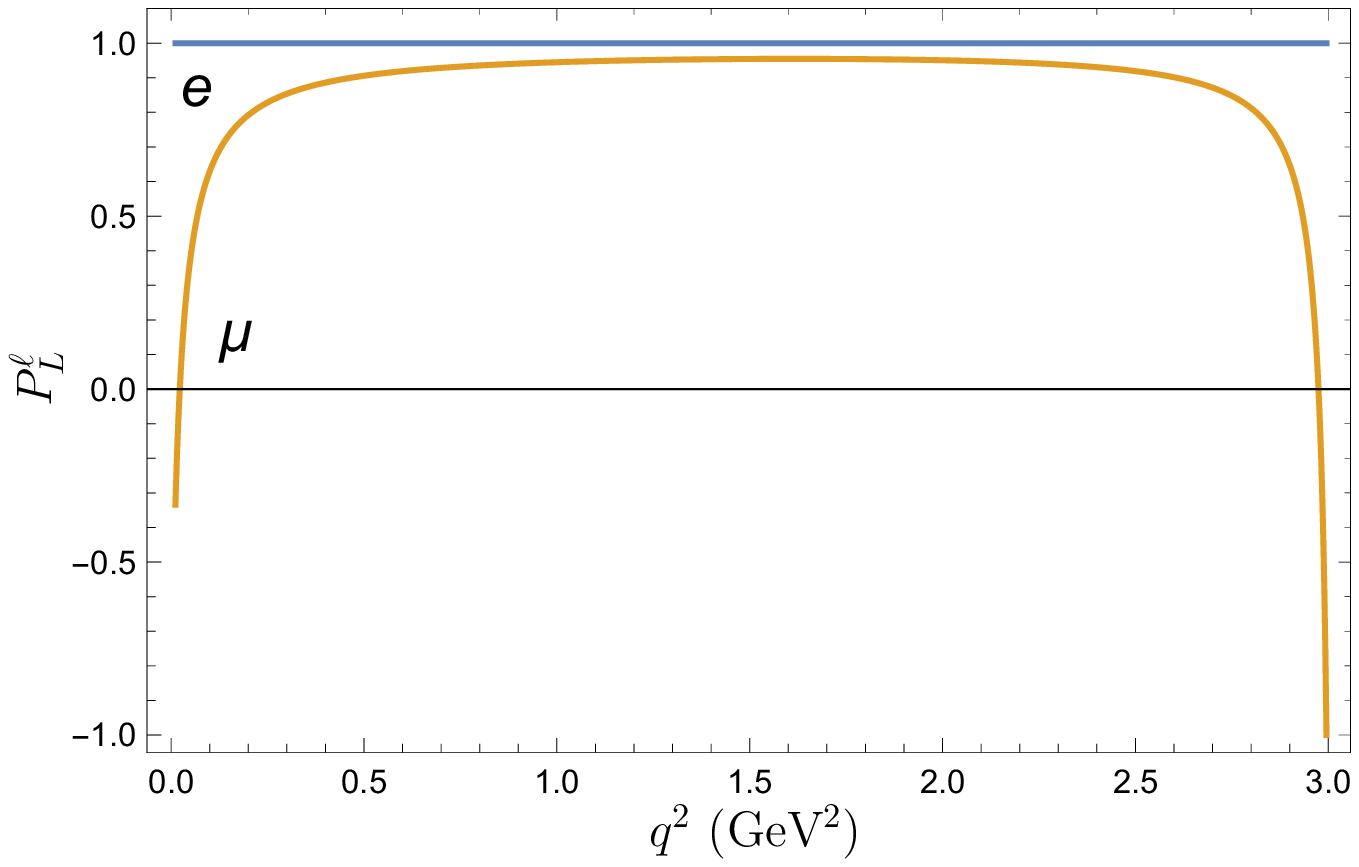}\ \
  \ \includegraphics[width=8cm]{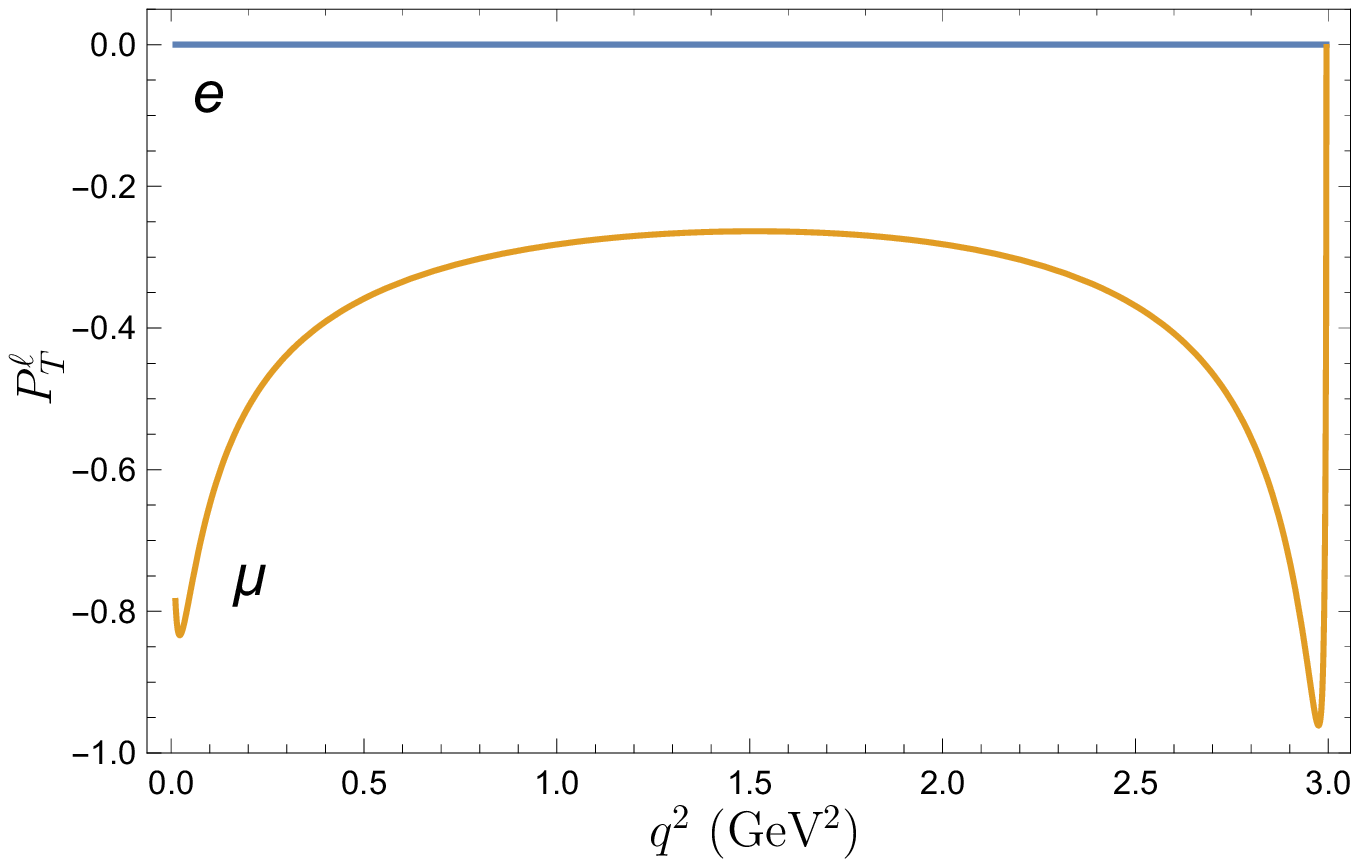}\\
\caption{Polarization and asymmetry parameters for the semileptonic
  $D^+\to \pi^{0}\ell^+\nu_\ell$ decays.    }
\label{fig:PApi}
\end{figure}

\begin{figure}
\centering
  \includegraphics[width=8cm]{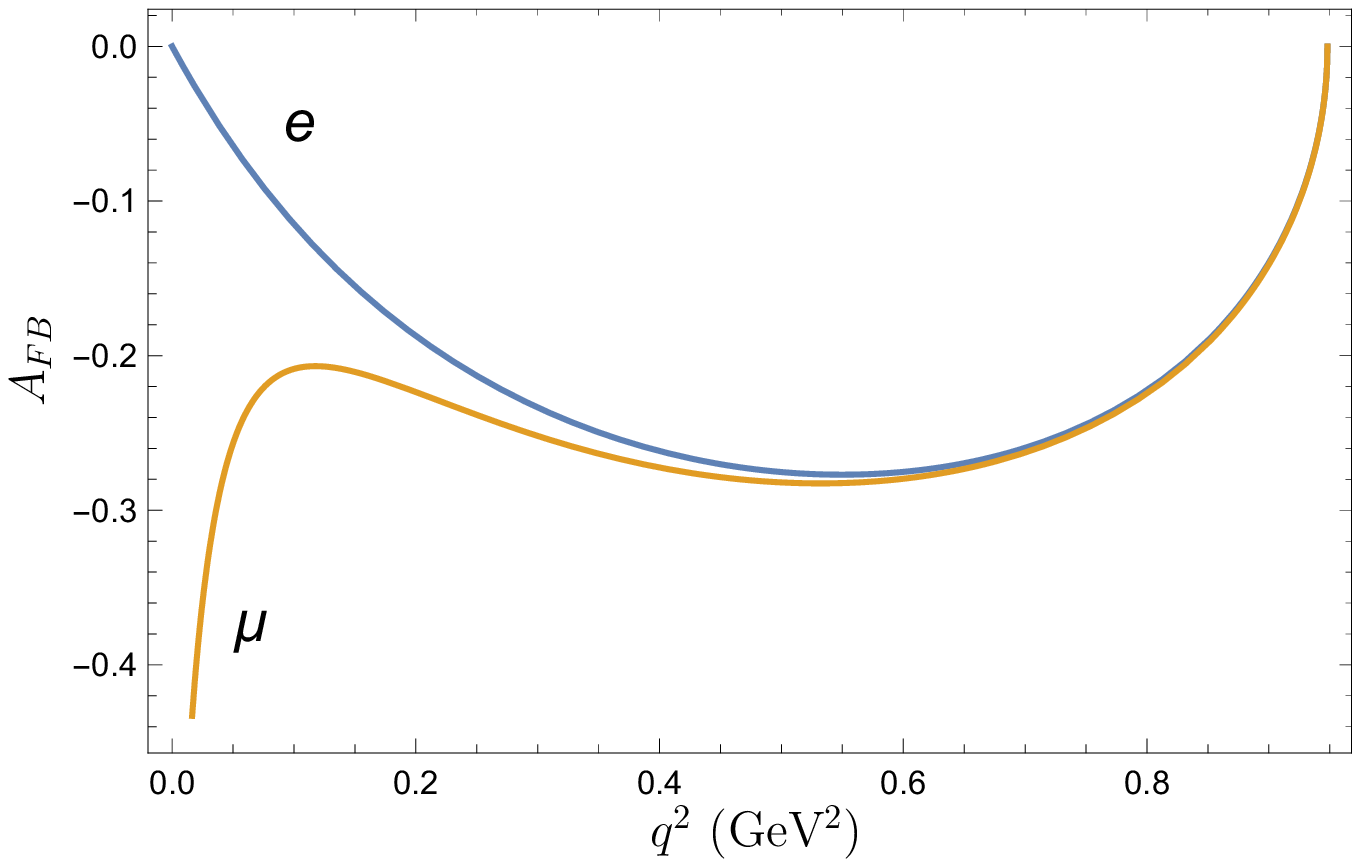}\ \
  \ \includegraphics[width=8cm]{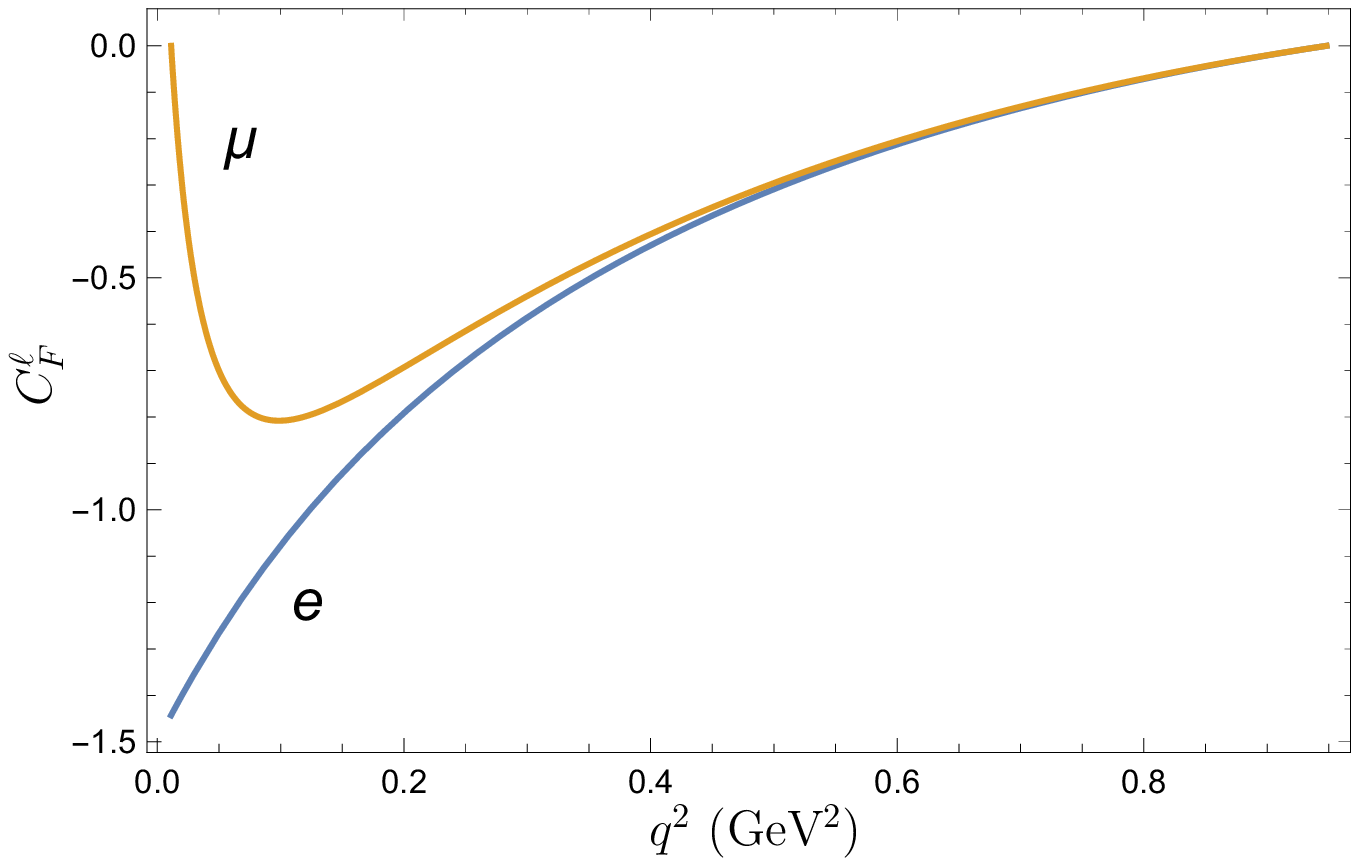}\\
  \includegraphics[width=8cm]{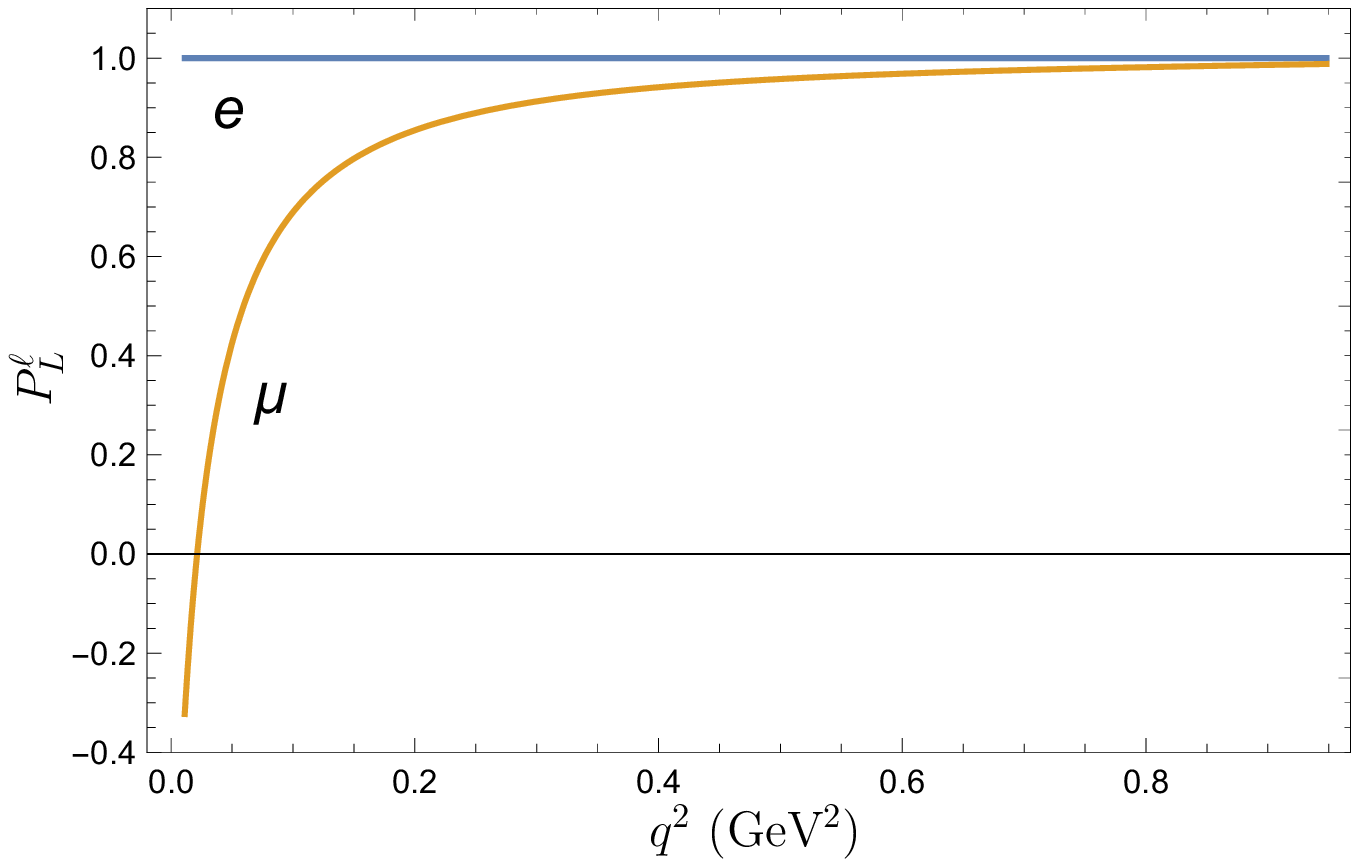}\ \
  \ \includegraphics[width=8cm]{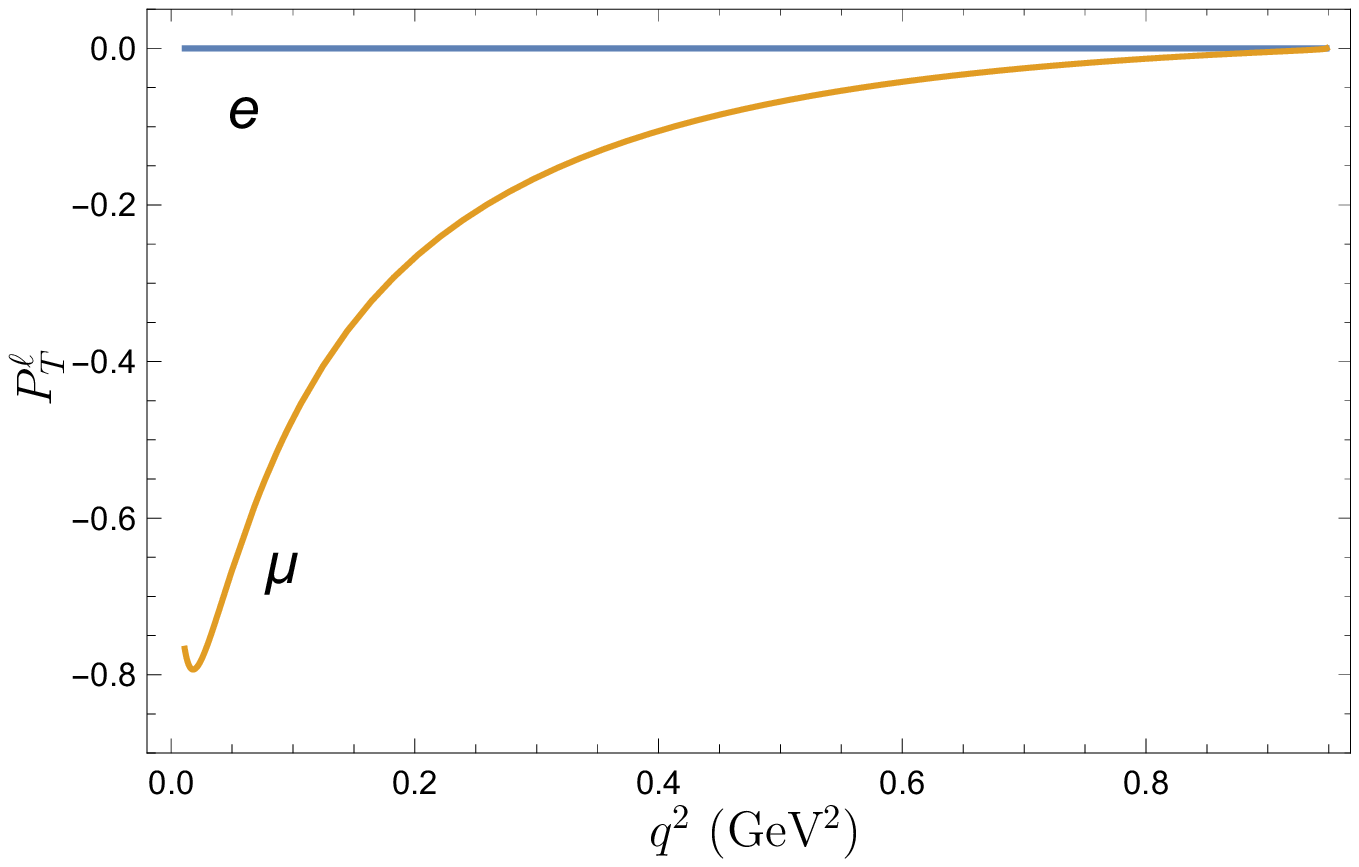}\\
\caption{Polarization and asymmetry parameters for the semileptonic
  $D^+\to \bar K^{*0}\ell^+\nu_\ell$ decays.    }
\label{fig:PA}
\end{figure}

We also calculate the forward-backward asymmetry $A_{FB}(q^2)$
[Eq.~(\ref{eq:afb})], the lepton-side convexity parameter $C_F^\ell(q^2)$ [Eq.~(\ref{eq:clf})],
the longitudinal $P_L^\ell(q^2)$ [Eq.~(\ref{eq:ple})] and transverse
$P_T^\ell(q^2)$ [Eq.~(\ref{eq:pte})]
polarization of the final charged lepton, and longitudinal
polarization $F_L(q^2)$ of the final vector meson, Eq.~(\ref{eq:fl}). As an example in
Figs.~\ref{fig:PApi} and \ref{fig:PA} we plot these asymmetries and
polarization parameters for $D^+\to\pi^0\ell^+\nu_\ell$ and
$D^+\to\bar K^{*0}\ell^+\nu_\ell$ decays.

In Table~\ref{pa} we present our predictions for the mean values
of the the polarization and asymmetry parameters for the semileptonic
$D$ and $D_s$ decays. These values were obtained by separately
integrating corresponding partial differential decay rates in
numerators and the total decay rates in denominators. Since we neglect
the small positron mass, for all decays  $D_{(s)}^+\to F e^+\nu_e$, $\langle
P^e_{L}\rangle=1$ and $\langle P^e_{T}\rangle=0$, while for
decays $D_{(s)}^+\to P e^+\nu_e$,  $\langle A_{FB}\rangle=0$
and $\langle C^e_{F}\rangle=-1.5$. Note that in Ref.~\cite{ikpsst}
close values of these parameters were found. Experimentally only the
ratios of the partial decay rates of the final vector meson states with
longitudinal and transverse polarization $\Gamma_L/\Gamma_T=\langle F_{L}\rangle
/(1-\langle F_{L}\rangle)$ have been measured for  $D^+\to \bar
K^{*0}\ell^+\nu_\ell$ and $D_s\to \phi \ell^+\nu_\ell$ decays. The
experimental  values for these ratios are  $1.13\pm0.08$ and
$0.72\pm0.18$ \cite{pdg}, respectively. The first value is in
agreement with our prediction $\Gamma_L/\Gamma_T=1.16$, while the second
one is somewhat smaller than predicted $\Gamma_L/\Gamma_T=1.19$.
Values of these ratios, close
to ours, were obtained in Ref.~\cite{ikpsst}.

\begin{table}
\caption{Predictions for the polarization and asymmetry parameters for the semileptonic
  $D$ and $D_s$ decays. }
\label{pa}
\begin{ruledtabular}
\begin{tabular}{cccccc}
Decay&$\langle A_{FB}\rangle$  & $\langle C^\ell_{F}\rangle$& $\langle P^\ell_{L}\rangle$  &$\langle P^\ell_{T}\rangle$& $\langle F_{L}\rangle$ \\
  \hline
  $D_{(s)}^+\to P e^+\nu_e$ &$0$ & $-1.5$ & $1$ &$0$ &\\
   $D^+\to \bar K\mu^+\nu_\mu$ &$-0.053$ & $-1.34$ & $0.85$ &$-0.42$ &
  \\
   $D^+\to \pi^0\mu^+\nu_\mu$ &$-0.040$ & $-1.38$ & $0.89$ &$-0.36$ &
  \\
   $D^+\to \eta\mu^+\nu_\mu$ &$-0.052$ & $-1.34$ & $0.85$ &$-0.40$ &
  \\
  $D^+\to \eta'\mu^+\nu_\mu$ &$-0.097$ & $-1.20$ & $0.72$ &$-0.56$ & \\
  $D^+\to \bar K^{*0}e^+\nu_e$ &$-0.22$ & $-0.47$ & $1$ &0 &0.54 \\
  $D^+\to \bar K^{*0}\mu^+\nu_\mu$ &$-0.25$ & $-0.37$ & $0.90$ &$-0.15$ &0.54 \\
  $D^+\to \rho^0e^+\nu_e$ &$-0.26$ & $-0.42$ & $1$ &0 &0.52 \\
  $D^+\to \rho^0\mu^+\nu_\mu$ &$-0.28$ & $-0.34$ & $0.92$ &$-0.12$ &0.52 \\
   $D^+\to \omega e^+\nu_e$ &$-0.25$ & $-0.39$ & $1$ &0 &0.51 \\
  $D^+\to \omega \mu^+\nu_\mu$ &$-0.27$ & $-0.32$ & $0.93$ &$-0.11$ &0.50 \\
$D_s\to \bar K^{0}\mu^+\nu_\mu$ &$-0.038$ & $-1.38$ & $0.89$ &$-0.34$ & \\
 $D_s^+\to \eta\mu^+\nu_\mu$ &$-0.043$ & $-1.37$ & $0.88$ &$-0.35$ &
  \\
  $D_s^+\to \eta'\mu^+\nu_\mu$ &$-0.080$ & $-1.26$ & $0.77$ &$-0.51$ & \\
  $D_s\to \bar K^{*0}e^+\nu_e$ &$-0.26$ & $-0.41$ & $1$ &0 &0.52 \\
  $D_s\to \bar K^{*0}\mu^+\nu_\mu$ &$-0.29$ & $-0.33$ & $0.92$ &$-0.11$ &0.51 \\
  $D_s\to \phi e^+\nu_e$ &$-0.21$ & $-0.49$ & $1$ &0 &0.54 \\
  $D_s\to \phi\mu^+\nu_\mu$ &$-0.24$ & $-0.35$ & $0.90$ &$-0.15$ &0.54 \\
\end{tabular}
\end{ruledtabular}
\end{table}

\section{Conclusions}
\label{sec:conc}
In the framework of the relativistic quark model based on the quasipotential
approach  we calculated the form factors of the semileptonic
$D$ and $D_s$ meson transitions. The relativistic effects including wave function
transformations from the rest to moving reference frame and contributions
of the intermediate negative energy states were consistently taken
into account. This allowed us to reliably calculate the form factors
in the whole accessible kinematical range without additional
approximations and/or extrapolations. The form factors were
expressed through the overlap integrals of the meson wave functions. These
wave functions were obtained in our previous study of the heavy-light
and light meson spectroscopy. This fact significantly increases
self-consistency and reliability of our approach, since in most of the
previous quark model studies of semileptonic decays some ad hoc
form of the wave function (mostly Gaussian) had been used.  It was
found that our numerical
results for the form factors and their $q^2$ dependence can be well
approximated by Eqs. (\ref{fitfv}) and  (\ref{fita12}). The parameters of
the fit are collected in Tables~\ref{ffD},~\ref{ffDs}. The calculated values of
the form factors $f_+(q^2)$ for the decays to pseudoscalar mesons and the ratios
of the form factors $r_2$ and $r_V$ for the decays to vector mesons  at $q^2=0$ agree
within errors with the experimental data (see Tables~\ref{ffcomp},~\ref{rcomp}).  The form factors $f_+(q^2)$ of the
weak $D$  transitions to pseudoscalar $K$ and $\pi$ mesons agree well with data form
Belle \cite{belleff}  and BaBar \cite{babarff} Collaborations in the whole $q^2$ range.

These form factors were applied for the calculation of  differential
and total decay rates of semileptonic decays of $D$ and $D_s$ using the
helicity formalism. The obtained differential decay distributions
(\ref{eq:dg}) for $D$ decays to pseudoscalar mesons, plotted in
Fig.~\ref{fig:GDcomp},  agree  with experimental data from the  BaBar
\cite{babarff} and  CLEO \cite{cleo} Collaborations. The calculated
total branching fractions are also in good agreement with averaged
experimental values from PDG \cite{pdg} and recent data from the BES
III Collaboration \cite{besiii-1,besiii-2,besiii-3,besiii-4,besiii-5,besiii-6,besiii-7,besiii-8,besiii-9}. To test lepton universality in the
semileptonic $D_{(s)}$ meson decays we calculated the ratios of the
branching fractions of decays involving muons to the ones involving
positrons. These ratios are collected in Table~\ref{lut} in comparison
with other theoretical predictions and available experimental
data. Within current experimental accuracy no deviations of data from the
standard model predictions are found. We also calculated the forward-backward
asymmetries, the lepton and vector meson longitudinal and transverse
polarization parameters which can be measured in future
experiments. Their mean values are collected in Table~\ref{pa}.

We presented the detailed comparison of our results with other
theoretical predictions. In most cases better agreement of our values
with data is found.  The further increase of the experimental accuracy
and new measurements can help to better understand quark dynamics in mesons.
This could be realized in the future super tau-charm facility \cite{Bondar,IPAC2018,PengCharm2018} which is hotly discussed to construct.

We are grateful to D. Ebert and M. Ivanov for valuable
discussions. This work was supported in part by
the National Natural Science Foundation of
China under Project Nos.11805012 and U1832121.

\end{document}